\documentclass{aa}  

\usepackage{graphicx}
\usepackage{gensymb}
\usepackage{natbib}
\usepackage{diagbox}  
\usepackage{siunitx}
\bibpunct{(}{)}{;}{a}{}{,}
\usepackage[colorlinks=true, allcolors=blue]{hyperref}
\usepackage{txfonts}

\titlerunning{Dwarf-Dwarf interactions and their influence on star formation}
\authorrunning{Chauhan et al.}
\begin{document} 

\title{Dwarf-Dwarf interactions and their influence on star formation: Insights from post-merger galaxies}

   \author{ Rakshit Chauhan
          \inst{1,2},
          Smitha Subramanian\inst{1,3}, Deepak A. Kudari\inst{1,4}, S. Amrutha\inst{1,2}, Mousumi Das\inst{1} 
          }

   \institute{Indian Institute of Astrophysics, Koramangala II Block, Bangalore-560034, India \\
              \email{rakshit.chauhan@iiap.res.in}
         \and
            Pondicherry University, R.V. Nagar, Kalapet, 605014, Puducherry, India 
            \and
            Leibniz-Institut fur Astrophysik Potsdam (AIP), An der Sternwarte 16, D-14482 Potsdam, Germany
            \and
            Indian Institute of Science Education and Research, Mohanpur, Haringhata Farm, West Bengal-741246, India }

  \date{Received XXXX; accepted XXXX}

\abstract 
    {Interactions and mergers play a crucial role in shaping the physical properties of galaxies. 
    Dwarf galaxies are the dominant galaxy population at all redshifts, and the majority of mergers are expected to occur between them. The effect of dwarf-dwarf mergers on star formation in these systems is not yet fully understood. In this context, we study the star formation properties of a sample of 6,155 isolated (i.e., with no massive galaxy, M$_{*}$ $>$ 10$^{10}$ M$_{\odot}$, within a 1 Mpc$^3$ volume) dwarf galaxies consisting of 194 post-merger and 5,961 non-interacting sample, spanning a stellar mass range of $ 10^{7} - 10^{9.6} M_{\odot} $ and a redshift range of 0.01 -- 0.12. 
    The post-merger galaxies studied here were identified in a past study in the literature, which found galaxies with signatures of recent merger activity (in the form of tidal features) in deep optical images. We use the Far Ultraviolet imaging data from the GALEX mission and estimate the star formation rate (SFR) of our sample galaxies. To investigate the impact of interactions on star formation, we estimated the difference in log(SFR) between a post-merger galaxy and the median of its corresponding control sample matched in stellar mass and redshift. The offset in our sample has a range ($-$2 to $+$2 dex), indicating both enhancement and suppression of star formation in these recent merger galaxies. Around 67\% of the sample (130 galaxies) show an enhancement in SFR. The median offset (enhancement) of the sample is 0.24 dex (1.73 times), indicating $\sim$ 70\% increase in the SFR of recent merger galaxies compared to their non-interacting counterparts. Out of 194 post-merger dwarfs, around 44\%, 20\%, and 9\% show 2, 5, and 10 times enhancements in SFR, respectively. Overall, we found a moderate enhancement in the median SFR of the post-merger sample, compared to that of the non-interacting dwarfs, by a factor of nearly two. This factor is comparable to the average enhancement factor observed in massive post-merger galaxies. However, we observed widespread star formation across the sample of dwarf galaxies. Star formation is found to be enhanced in both the central (6" diameter region at the centre) and outer regions of the post-merger galaxies compared to their non-interacting counterparts, and the factor of enhancement was found to be similar. This is in contrast to what is observed in massive galaxies, where the merger-triggered star formation is more significant in the central regions. Furthermore, we didn't observe any significant dependence of the enhancement factor on stellar mass across the sample. Additionally, we found that in the given small range of redshift, post-merger dwarfs exhibit a higher median specific star formation rate compared to their non-interacting counterparts. About 33\% of the galaxies in our post-merger dwarf sample are quenched. These galaxies could be at a later stage of the post-merger regime, where quenching can happen as observed in massive galaxies. This study suggests that dwarf-dwarf mergers can affect star formation in the local universe. A more comprehensive study of post-merger dwarfs is required to understand their evolution.}


   \maketitle
%
\section{Introduction}
Dwarf galaxies dominate the galaxy population at all redshifts \citep{1988ARA&A..26..509B,2006A&A...459..745F,2015A&A...575A..96G} and play a significant role in the hierarchical assembly of galaxies. Most mergers are predicted to occur among dwarf galaxies, with an average of three major mergers during their evolution  \citep{2010MNRAS.406.2267F}. Therefore, it is essential to understand the role of dwarf-dwarf mergers in their evolution. While there exists a plethora of studies about the effect of massive ($M_{*}$ $>$ 10$^{10}$ M$_{\odot}$) galaxy mergers on star formation (\citealt{10.1111/j.1365-2966.2010.17076.x,2011MNRAS.418.2043E,2011MNRAS.412..591P,2013MNRAS.433L..59P,2016MNRAS.461.2589P,2020MNRAS.494.4969P,2024MNRAS.529.1493P,2013MNRAS.430.1901H,2016MNRAS.462.4495H, Shah_2022, Brown_2023,2023ApJ...953...91L,2025MNRAS.538L..31F} and references therein), the effects of dwarf-dwarf interactions remain less explored. Dwarf galaxies have shallow gravitational potential and are more affected by baryonic feedback and environmental effects (\citealt{2010MNRAS.402.1599S,2019A&A...625A.143V,2021MNRAS.503..176H,2023A&A...670A..92M} and references therein), as compared to massive galaxies. As a result, it remains uncertain whether dwarf-dwarf interactions can initiate star formation or induce morphological changes, as observed in interacting massive galaxies. Also, due to the very steep and uncertain relation between galaxy mass and dark matter halo mass, it is unclear whether the bottom-up accretion-driven formation process predicted by $\Lambda$ cold dark matter (CDM) can be observed through tidal debris and stellar halos at the scale of dwarf galaxies \citep{2022MNRAS.511.4044D}.

Observational studies face significant challenges in studying dwarf galaxies as they often have low luminosities \citep{2002AJ....124.2600W,2003MNRAS.339...33H,2007AJ....133..715W} and complex substructures surrounding them. On the theoretical side, it is often challenging to achieve the necessary resolution to accurately model dwarf galaxies along with sufficiently large volumes to establish a realistic cosmological context \citep{2014MNRAS.444.1518V,2017ARA&A..55...59N}. Most of the previous observational studies, which explored the role of interactions on their evolution, focused on individual systems \citep{2012ApJ...748L..24M,2015AJ....149..114P,2017ApJ...846...74P,2017MNRAS.467.2980S,2021MNRAS.500.2757O,2022AJ....164...82P,2022A&A...666A.103S}. However, recent deep optical surveys \citep{2016ApJ...828L...5C,2016MNRAS.458.1678H,2020MNRAS.491.5101A} dedicated to study the assembly process on smaller scales, have identified many dwarf-dwarf pairs and studied the effect of these interactions on their star formation, Active Galactic Nuclei (AGN) activity, and morphological features.
Improved cosmological simulations \citep{2021MNRAS.500.4937M} have enabled us to study dwarf-dwarf interactions in greater detail.
 
From the study of 18 nearby starburst dwarf galaxies, \cite{10.1093/mnras/stu1804} found that starburst dwarfs have more asymmetric outer HI morphologies than typical dwarf irregulars and suggested that past interactions or mergers would have triggered the starburst activity. \cite{2015ApJ...805....2S} found an enhancement in the star formation rate (SFR) of 60 paired dwarfs (with a pair separation of, $R_{sep} \leq $ 50 kpc, mass ratio of the pair $<$ 10) by a factor of 2.3 $\pm$ 0.7, compared to that of the single dwarfs which are matched in stellar mass, redshift and environment (distance from a massive neighbor $>$ 1.5 Mpc). The enhancement was found to decrease with increasing pair separation and extended out to pair separations as large as 100 kpc. Their sample covered a mass range of $ 10^{7} - 10^{9.7} M_{\odot}$ (with a median mass of 10$^{8.9}$ M$_{\odot}$) and a redshift range of 0.005 $<$ z $<$ 0.07. Their study also found that close encounters ($R_{sep} \leq$ 50 kpc) between dwarf galaxies can enhance their SFR, irrespective of the presence of a massive galaxy within a distance of $<$ 1.5 Mpc. 
\cite{2020AJ....159..103K} observed a correlation between the existence of tidal features and star formation activity. \cite{2025ApJ...980..157H} investigated the SFR enhancement of 278 HI-rich galaxies in isolated galaxy pairs detected by the Widefield ASKAP L-band Legacy All-sky Blind surveY (WALLABY) with stellar mass 10$^{7.6} - 10^{11.2} M_{\odot}$ and redshift z $<$ 0.075. They found that the SFRs of interacting galaxies are enhanced compared to galaxies with similar stellar mass and HI mass, and the SFR enhancement increases with increasing pair proximity. However, they found that high mass  (stellar mass $>$ 10$^{9}$ M$_{\odot}$) and low mass galaxies (stellar mass $<$ 10$^{9}$ M$_{\odot}$) behave differently during the interactions. Tidal perturbations are more important in triggering star formation in high-mass galaxies than in low-mass galaxies. Recently, \cite{2025A&A...698A.260B} identified an isolated merging dwarf system, located in the centre of a cosmic void, using the Calar Alto Void Integral-field Treasury Survey (CAVITY, \citealt{2024A&A...689A.213P}) data. They found that the SFRs of the two components of the merging system are higher than those reported for similar mass star-forming dwarf galaxies and suggested that the merger could have contributed to enhancing star formation.

The optical integral field unit (IFU) observations of an interacting pair of dwarf galaxies, dm1647+21, using the Very Large Telescope/Multi Unit Spectroscopic Explorer (VLT/MUSE) observations showed that the star formation in this dwarf pair is not only enhanced compared to that of dwarf galaxies with similar mass but also widespread and clumpy \citep{2017ApJ...846...74P}. This spatial distribution of star formation contrasts with that observed in merging massive galaxies, where gas funnelling mainly leads to nuclear starburst. \cite{2020ApJ...894...57S} performed a spatially resolved study of star formation in low mass galaxies (with stellar masses in the range, 10$^{8} - 10^{10} M_{\odot}$ with a median stellar mass of  10$^{9.5} M_{\odot}$ with redshift $<$ 0.07) using the SDSS-IV/MaNGA IFU data. 
These authors observed an enhancement in the SFR of dwarf pairs (with pair separation of $<$ 100 kpc, the mass ratio of the pair of $<$ 4, and a line-of-sight kinematic separation of $\leq$ 100 kms$^{-1}$) in their inner regions, decreasing radially outwards. All these studies suggest that dwarf-dwarf interactions trigger star formation in the galaxies, and the spatial distribution of triggered star formation could be significantly different from what is observed in interacting massive galaxies. However, \cite{Paudel_2018} studied 177 interacting dwarfs having $M_{*} < 10^{10} M_{\odot}$ and z < 0.02 (with a median mass of $ 10^{9.1} M_{\odot} $ and median redshift of 0.01) and found no enhancement in star formation among interacting dwarfs compared to the local volume star-forming galaxies.

A recent study by \cite{2024ApJ...963...37K} suggests that dwarf-dwarf interaction could enhance as well as quench star formation. But considering the scarcity of quenched field dwarfs, they suggest that this quenching seems temporary (for a short period, $\sim$ 560 Myr). Additionally, due to their shallow gravitational potentials, dwarf galaxies are expected to be more prone to feedback from stars and AGN \citep{2025A&A...693L..16B}, which can affect their star formation. Dwarf galaxies are known to host AGNs (\citealt{2023ApJ...959..116N,2024ApJ...972L..24P}). A recent study by \cite{2024ApJ...968L..21M} found a statistically significant enhancement (4$\sigma$-–6$\sigma$) in AGN detection in an interacting sample of dwarf galaxies, compared to isolated dwarf galaxies. The authors suggest that the presence of a nearby dwarf neighbor plays a crucial role in triggering black hole accretion. \cite{2024MNRAS.532..613B} found lower star formation activity in dwarfs hosting AGN, but there was no signature of significant quenching in these dwarfs as suggested by simulations  \citep{10.1093/mnras/stz1616,2023ApJ...957...16S,arjonagalvez2024roleagnfeedbackevolution}. Thus, the impact of dwarf-dwarf interactions on their star formation remains inconclusive. 

Most of the previous studies, which focused on a relatively large sample of dwarf-dwarf interactions and their impact on their star formation, analyzed galaxies with stellar masses of the order of $ \sim10^{9} M_{\odot}$, representing the upper end of the low-mass regime. Additionally, most of these studies are based on observations at optical wavelengths. 
As young and massive O and B-type stars predominantly emit in the Ultraviolet \citep{1987A&A...180...12D,1998ARA&A..36..189K}, it is an optimal wavelength to study the star-forming properties of galaxies. Recently, \cite{2024A&A...681A...8S} studied dwarf galaxies, with stellar mass: $10^{6} - 10^{9} M_{\odot}$ (median mass of $\sim$ $10^{7.5} M_{\odot}$), using the Far-Ultraviolet (FUV) data from the GALEX mission. The authors examined a sample of 58 galaxies (22 interacting and 36 single gas-rich dwarf galaxies) in the Lynx-Cancer void. They found an enhancement in SFR by a factor of $3.4 \pm 1.2$ for the interacting systems compared to isolated dwarf galaxies in the stellar mass range of $ 10^{7} - 10^{8} M_{\odot}$. Although this study suggests a direct correlation between galaxy interactions and enhanced star formation, the sample size of dwarf galaxies in this study was relatively small and included dwarfs undergoing various types of interactions such as mergers, fly-bys, and pairs, which may introduce heterogeneity in the observed effects as suggested by \cite{2021MNRAS.500.4937M}.
Additionally, the study was restricted to dwarfs within a distance of $\sim25$Mpc. A recent study by \cite{2024ApJ...976...83K} demonstrated that the star-forming sequence (SFR vs stellar mass) of low-mass galaxies evolves significantly between 0.05 $<$ z $<$ 0.21. It would be interesting to explore the effect of dwarf-dwarf interactions on this redshift evolution of the star-forming main sequence of low-mass galaxies. 

One of the studies that analysed a large sample of dwarf galaxies to understand the effect of dwarf-dwarf mergers on their star formation is by  \cite{2020AJ....159..103K}. They identified a sample of 226 dwarf galaxies (from a sample of 6875 dwarf galaxies with stellar mass range of 10$^{7}$ -- 10$^{9.6}$ M$_{\odot}$ and in the redshift range of 0.01 -- 0.12) with signatures of recent merger activity. They found that the fraction of galaxies that host detectable tidal features increases strongly as a function of star formation
activity. They used the H$\alpha$  equivalent width and (g-i) colour of host galaxies as tracers of star formation activity. Though \cite{2020AJ....159..103K} found a dependence of tidal feature detection fraction on star formation, they did not measure and compare the SFRs of the merger and non-merger samples. 
Again, the H$\alpha$ equivalent width measurements in their study were from the central regions of the sample galaxies, specifically from the 2" -- 3" regions of these galaxies ($\sim$ 5 -- 8 kpc at the redshift of z = 0.12) using the fiber-based spectra from the Sloan Digital Sky Survey (SDSS) spectroscopic surveys or from the Galaxy and Mass Assembly (GAMA) spectroscopic survey. However, considering the spatial extent ($\sim$ 20 -- 30 kpc) of these dwarfs, spectra obtained solely from the central region may not provide a comprehensive view of star formation across the entire galaxy.  

\indent In this context, we estimate the instantaneous/current SFR of the sample provided by \cite{2020AJ....159..103K}, and compare the star formation properties of the merging and non-merging sample (matched in stellar mass and redshift). The effect of final coalescence and post-merger on the star formation of dwarf galaxies is not very well explored, and it is an ideal sample to study and quantify the same. We use the FUV archival science-ready imaging data from the GALEX mission. We also compare the SFR in these galaxies' central and outer regions to understand the effect of dwarf-dwarf interactions on the spatial distribution of star formation. As this sample has a relatively large number of galaxies, we also attempt to understand the redshift evolution of star formation in dwarf galaxies. 

     \begin{figure*}
     \centering
   \includegraphics[width=18.6cm]{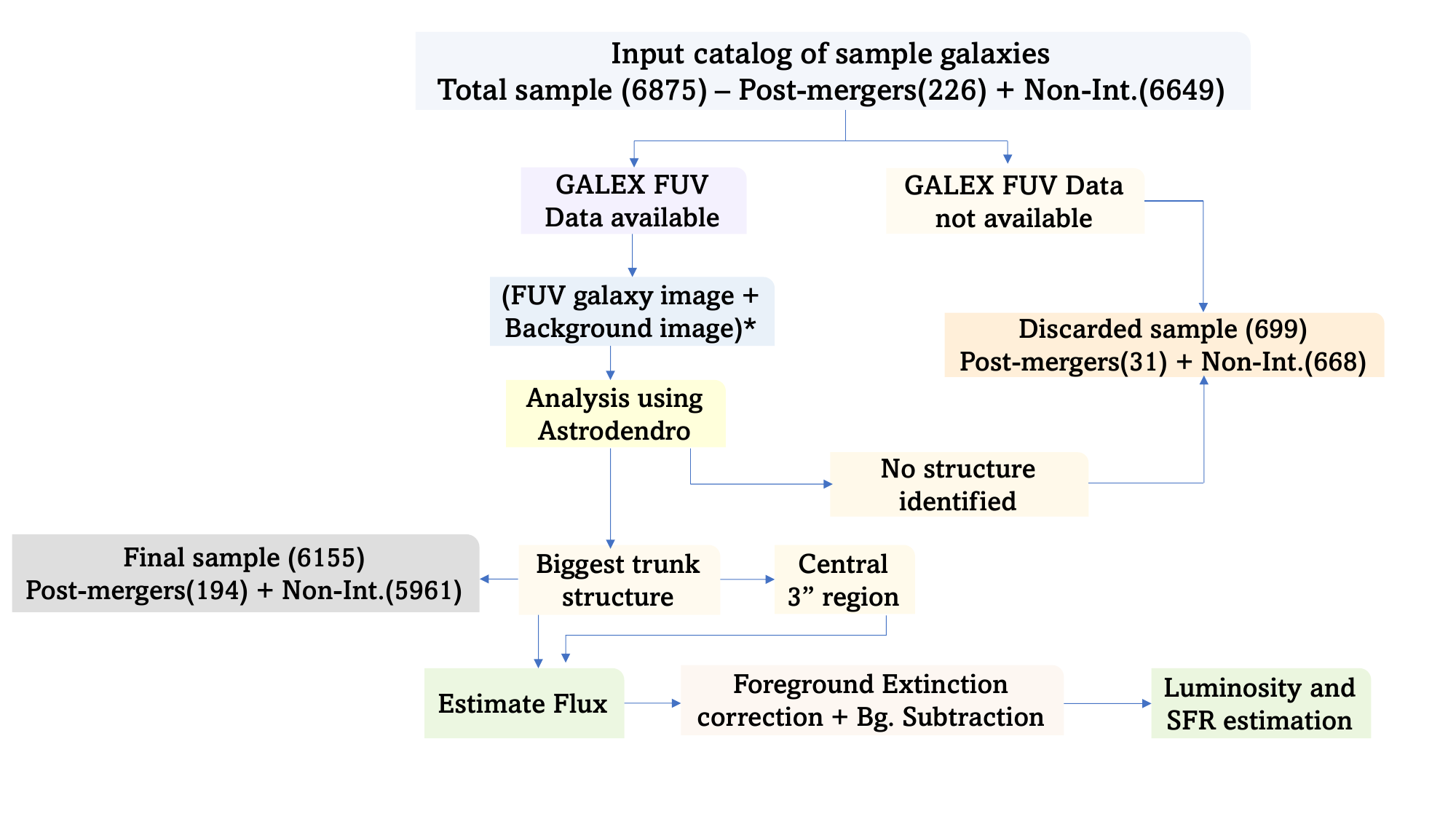}
   \caption{ Flowchart showing the automated analysis of sample dwarfs step by step. For available data, we downloaded the FUV image containing the dwarf galaxy along with the corresponding background image. (*)-If multiple FUV datasets are available, then we download the image having maximum exposure. The subsequent analysis steps are detailed in Section \ref{3}.}
              \label{Fig1}%
    \end{figure*} 

The paper is structured as follows: The sample selection is presented in Section \ref{2}, data and analysis are outlined in Section \ref{3}, followed by results and discussion in Sections \ref{4} \& \ref{5}. The summary of the study is presented in Section \ref{6}. Throughout this paper we assume the standard $\Lambda$CDM cosmology model with parameters h = H$_0$/(100 km$s^{-1}Mpc^{-1}$) = 0.6774, Ω$_m$ = 0.2589, Ω$_ \Lambda$ = 0.691, Ω$_b$ = 0.0486 \citep{2016A&A...594A..13P}.

\section{Sample selection} \label{2}    
For this study, we used the catalog of dwarf galaxies provided by 
\cite{2020AJ....159..103K}. It contains a sample of 6875 spectroscopically confirmed (z $<$ 0.12) isolated dwarf galaxies (with no massive galaxy, M$_{*}$ $>$ 10$^{10}$ M$_{\odot}$, within a 1 Mpc$^3$ volume), with 226 of them showing signatures of recent merger activity. They used high resolution imaging data obtained as part of the Hyper Suprime-cam Subaru Strategic Program (HSC-SSP; \citealt {2018PASJ...70S...4A,2018PASJ...70S...8A,2018PASJ...70S...5B,2018PASJ...70S...7C}), using the Hyper Suprime-cam at the 8.2 m Subaru telescope. The HSC–SSP reaches a surface brightness limit of $\sim$ 27 mag $arcsec^{-2}$ and can detect isolated low surface brightness (LSB) structures \citep{2018ApJ...857..104G,2018ApJ...866..103K}. \cite{2020AJ....159..103K} used HSC S18A data release, which covers over 300 $deg^2$ on the sky in $g_{HSC}, r_{HSC}$, and $i_{HSC}$ bands, and the updated version of an automated tidal feature detection algorithm, provided by \cite{2018ApJ...866..103K}, to detect LSB or tidal features. The automated tidal feature algorithm detected 226 isolated dwarf galaxies with LSB or tidal features. In visual inspection, 101 out of these 226 are found to host unambiguous signatures of dwarf-dwarf interactions. Based on the sensitivity limits of their data, they suggested that these samples of galaxies with tidal features are most likely to be the result of recent major mergers (with mass ratio of pair $>$ 1:4) in the dwarf regime. These are ideal candidates to study the effect of dwarf-dwarf mergers on star formation.

All the 6875 sample galaxies (226 with tidal features and 6649 without tidal features) in the catalog have spectroscopic observations either from the SDSS  spectroscopic surveys (both legacy and BOSS surveys, \citealt{2002AJ....124.1810S,2013AJ....145...10D,2016MNRAS.455.1553R}) or from the GAMA spectroscopic survey \citep{2018MNRAS.474.3875B}. The catalog provides the stellar mass and the H$\alpha$ emission line properties (equivalent width and total line flux) from the GAMA and SDSS spectroscopic databases, along with the spectroscopic redshifts. It also provides the $(g-i)$ colors of the sample galaxies. Stellar masses provided in the GAMA database \citep{2011MNRAS.418.1587T} are measured assuming a Chabrier initial mass function, IMF \citep{2003ApJ...586L.133C}. 
The stellar masses provided in the SDSS database are calculated using the flexible stellar population synthesis models of \cite{2009ApJ...699..486C}, assuming the Kroupa IMF \citep{2001MNRAS.322..231K}. Based on the stellar masses of the common sample in the SDSS and GAMA databases, \cite{2020AJ....159..103K} found that the SDSS-provided stellar masses are systematically higher than the equivalent measurement in the GAMA database, by a median of 0.08 dex. Hence, they reduced the masses derived from SDSS observations by 0.08 dex. These dwarf galaxies cover a stellar mass range of 10$^{7}$ -- 10$^{9.6}$ M$_{\odot}$. For our study, we consider all 226 galaxies identified by the automated tidal feature algorithm as the post-merger sample, and the remaining 6649 as the non-interacting sample.  
   \begin{figure*}
   \centering
   \includegraphics[width=19.1cm]{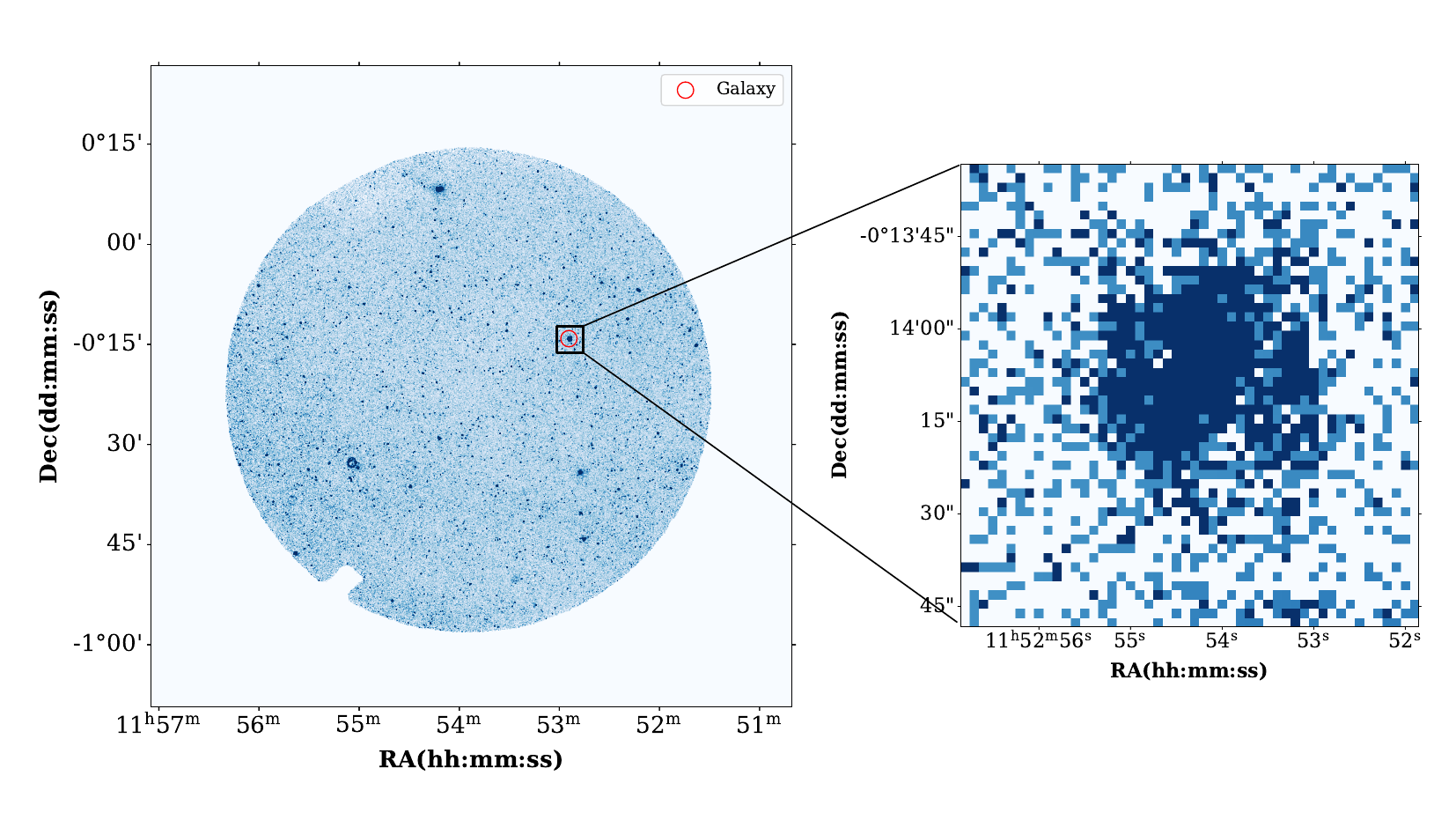}
   \caption{ Left panel image displays the entire field of view 1.2$^{\circ}$ of GALEX with a red circle highlighting the location of the dwarf galaxy LEDA 1148477. The right panel image shows a zoomed-in view of the dwarf galaxy.}
              \label{Fig2}%
    \end{figure*}

   \begin{figure*}
   \centering
   \includegraphics[width=9.0cm]{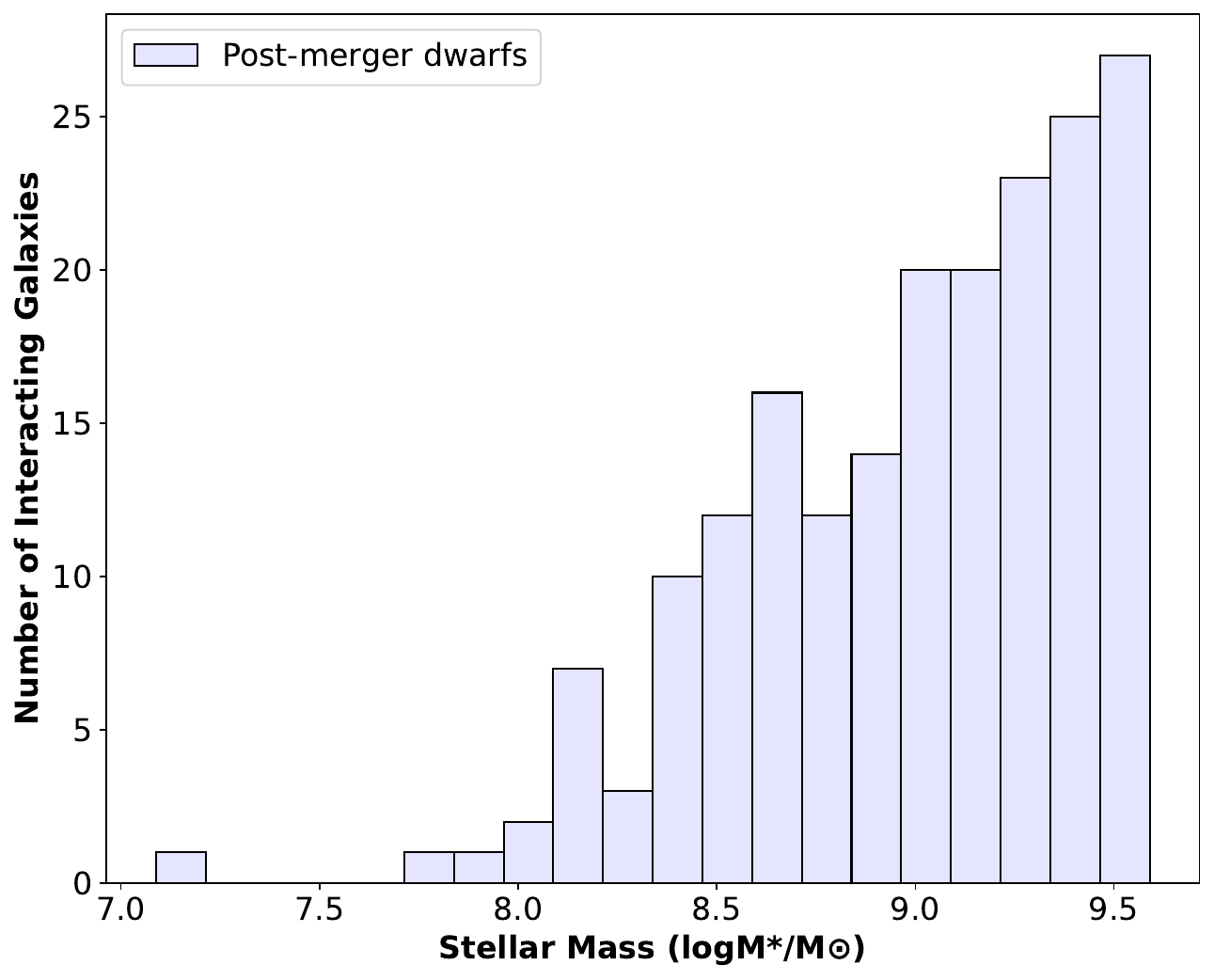}
   \includegraphics[width=9.0cm]{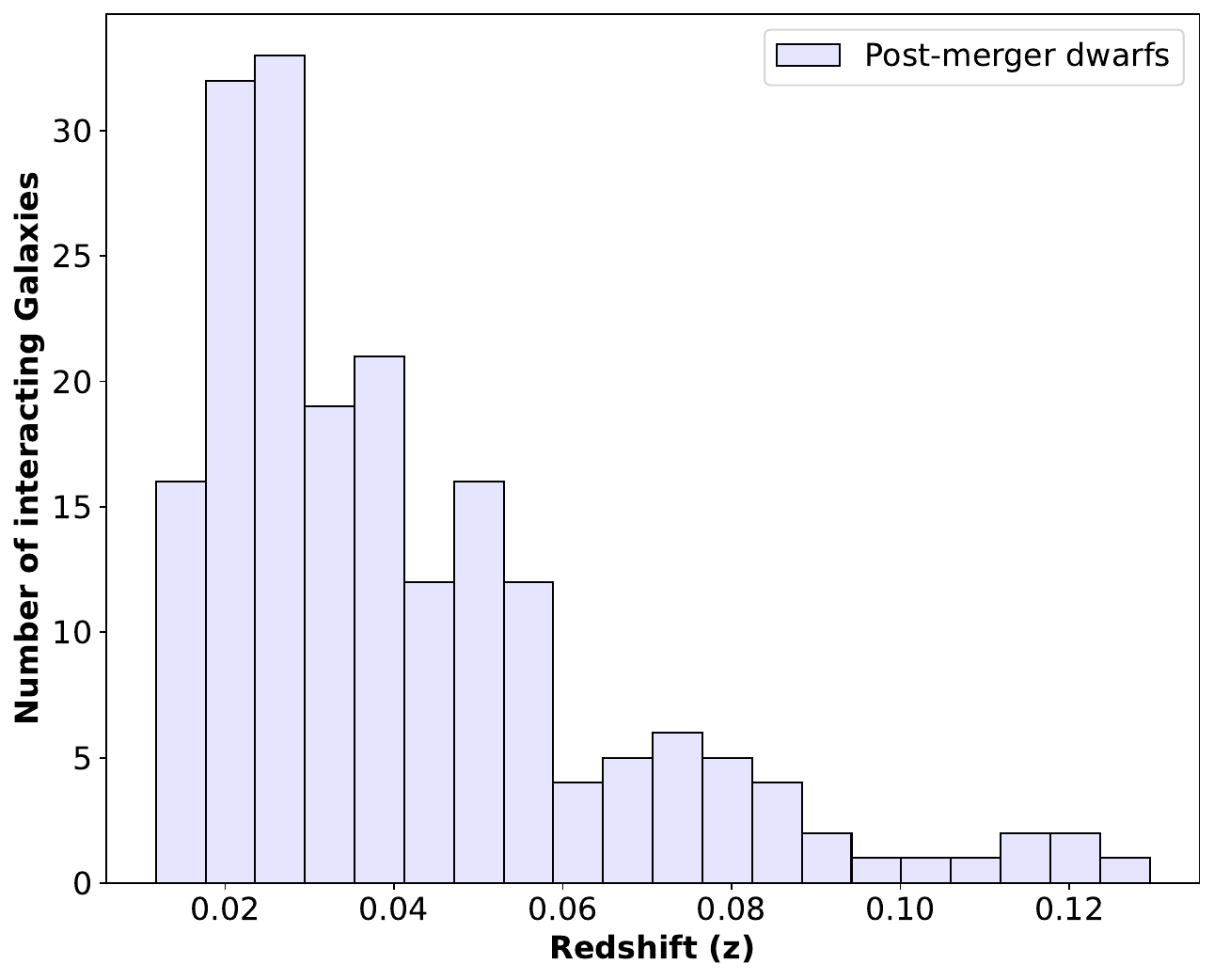}
   \includegraphics[width=9.0cm]{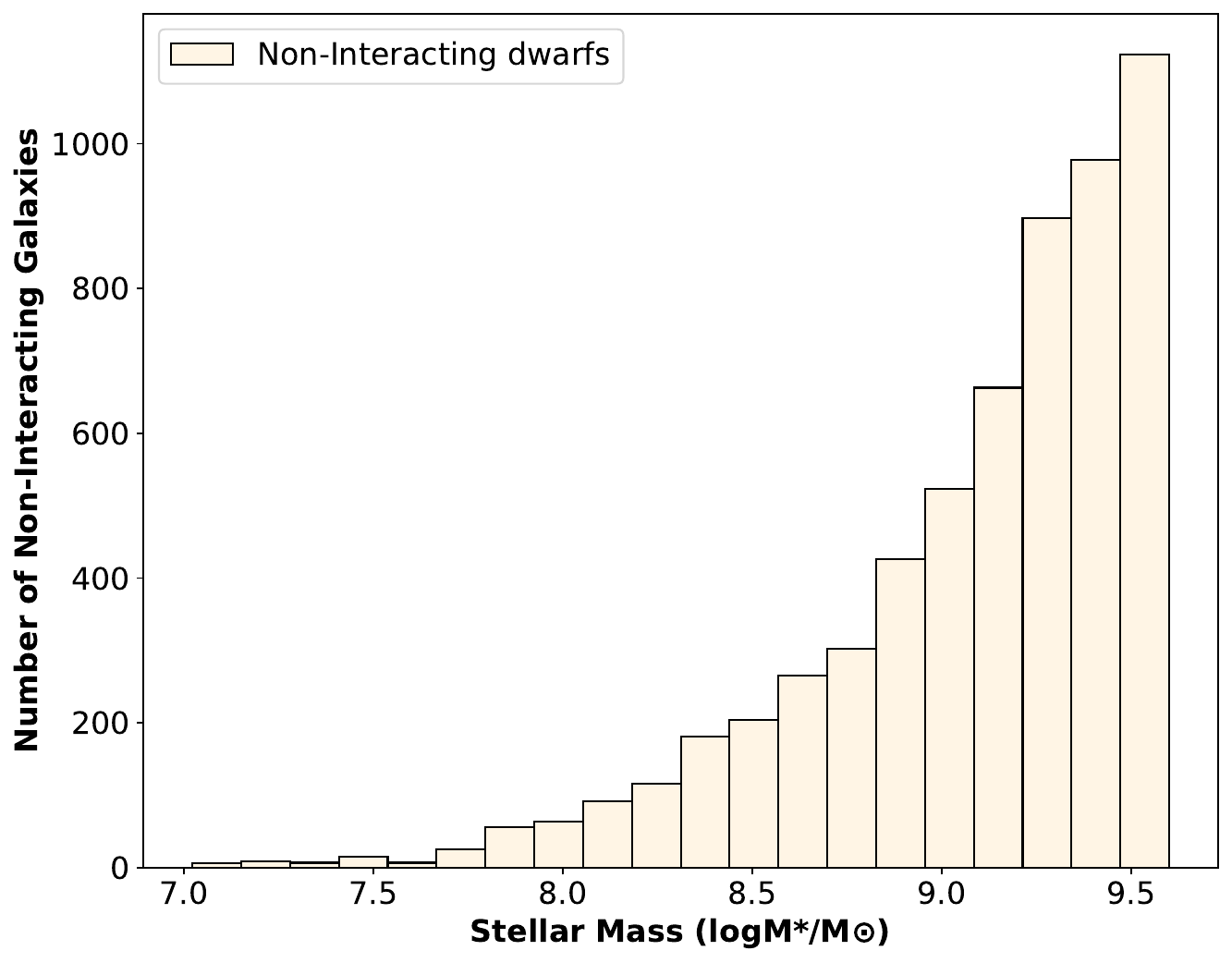}
   \includegraphics[width=9.0cm]{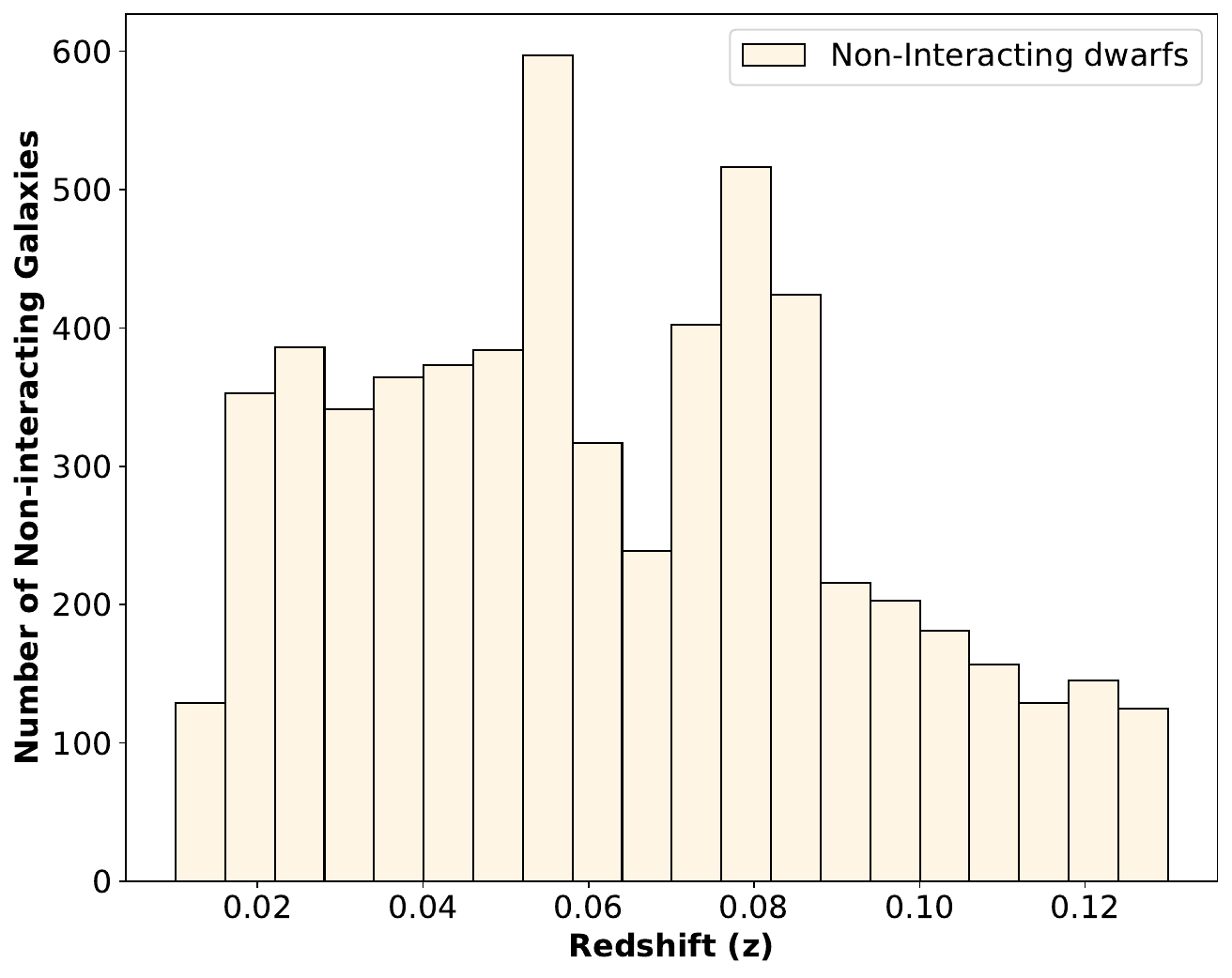}
   \caption{ Histogram displaying the distribution of target galaxies corresponding to stellar mass (top) and redshift (bottom). The blue bars represent post-mergers, while the orange bars represent the non-interacting dwarfs. Stellar mass is given in units of Solar mass. }
              \label{Fig3}%
    \end{figure*}
    
\section{Data and Analysis} \label{3} 
This study aims to estimate and compare the instantaneous/current SFR of post-merger and non-interacting isolated dwarf galaxies in our sample. Young star-forming regions, which host massive O and B-type stars, predominantly emit in UV, and hence, it is an optimal wavelength to investigate the star-forming properties of galaxies. The UV emission from a galaxy can trace star formation up to 100 -- 300 Myr \citep{1984A&A...140..325D,Schmitt_2006,2012ARA&A..50..531K}. FUV is particularly effective in tracing recent star formation in the last 100 Myr \citep{2015AJ....149...51C} and the current SFR of a galaxy is proportional to its FUV luminosity. 

In this study, we use the FUV ($\lambda_{eff} = 1538.6 {\AA}$) data from the NASA GALEX mission \citep{2005ApJ...619L...1M} to estimate the SFR of our sample galaxies. The GALEX observations were carried out using a Ritchey-Chretien-type telescope with a 0.5 m primary mirror and a focal length of 299.8 cm. Light was fed to the FUV and NUV detectors simultaneously using a dichroic beam splitter, and the instrument had an extensive $1.2^{\circ}$ field of view. GALEX conducted multiple surveys, including All-Sky, Medium, and Deep Imaging Surveys (shallower to deeper observations), a Nearby Galaxy Survey, and supported several Guest Investigator programs. 
The spatial resolution of FUV images is $\SI{4.2}{\arcsecond}$ \citep{2007ApJS..173..682M}. The calibrated science-ready FUV images can be obtained from the Mikulski Archive for Space Telescopes (MAST) \footnote{\url{https://mast.stsci.edu/portal/Mashup/Clients/Mast/Portal.html}} data archive. 
As we have a large sample of galaxies, we wrote a Python code to automatically download the GALEX FUV data of our sample galaxies from the MAST archive, and analyze them. The steps executed in the automated code are shown as a flowchart in Fig. \ref{Fig1}, and each of these steps is explained in the following sub-sections.

\textbf{\subsection{GALEX FUV data}} \label{3.1}
The Right Ascension (R.A.) and Declination (DEC.) of our sample galaxies, taken from the catalog by \cite{2020AJ....159..103K} are provided as input to the code. Based on the availability of the data in the MAST archive, the calibrated science-ready FUV images containing our sample galaxies and the corresponding sky background images are downloaded. If a sample galaxy has multiple GALEX FUV channel observations, the image corresponding to the highest exposure (exposure time range from $\sim$ 100s -- 100Ks, with the majority of galaxies having an exposure time of $\sim$ 2000s) is downloaded. Of 6875 galaxies, 6176 (195  post-merger dwarfs and 5981 non-interacting dwarfs) have GALEX FUV observations. For the remaining 699 dwarfs, either GALEX data were unavailable or had low exposure. We also removed those sources lying at the edges of the $1.2^{\circ}$ field of view image of GALEX.  Of 195 interacting dwarfs, classified based on the automated detection algorithm described in \cite{2018ApJ...866..103K}, 86 are visually confirmed interacting dwarfs. The GALEX FUV image of the full 1.2$^{\circ}$ field, containing one of our sample galaxies, is shown in Fig. \ref{Fig2}.
   \begin{figure}[ht!]
   \centering
   \includegraphics[width=9.0cm]{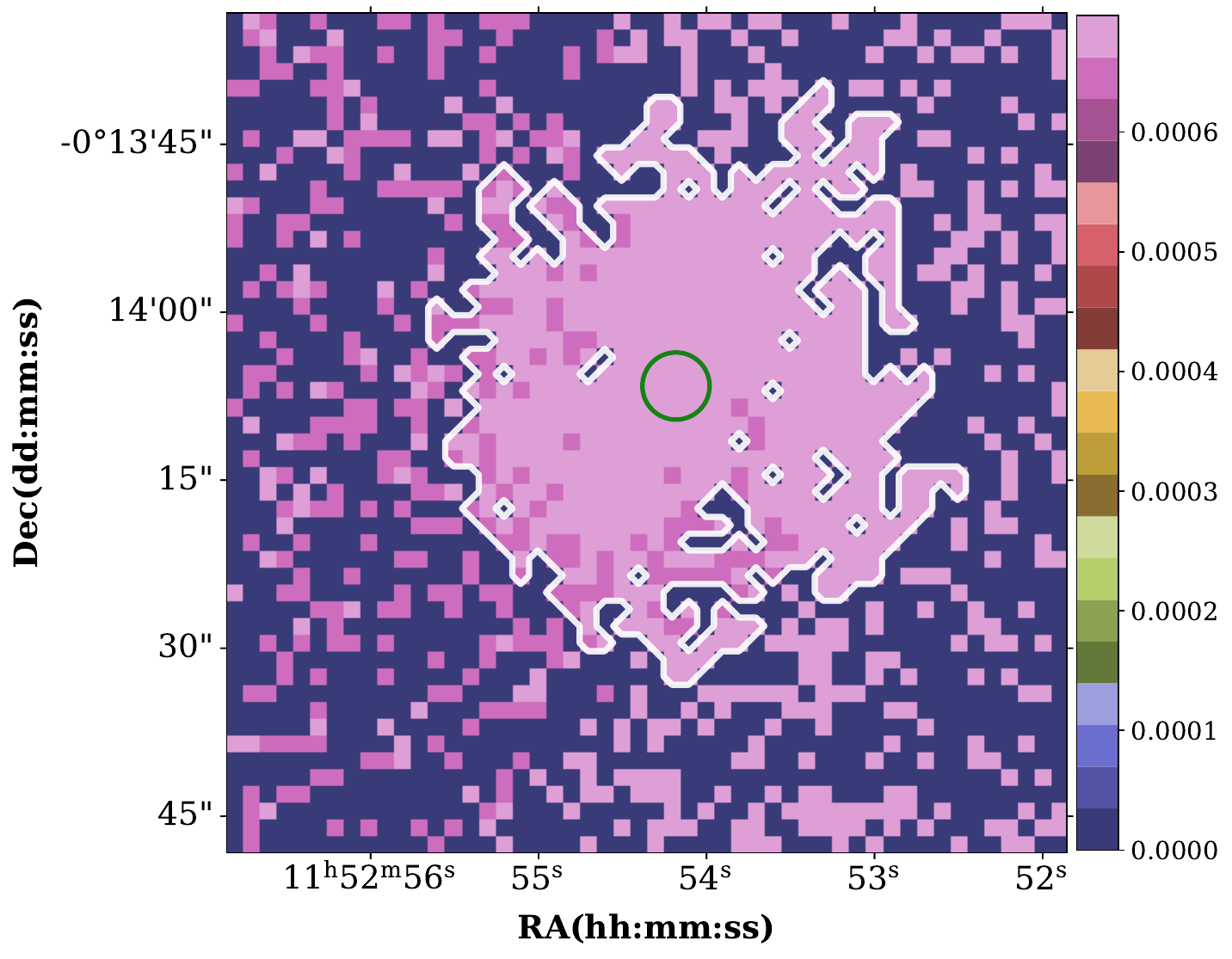}
      \caption{ Trunk structure (white boundary) of dwarf galaxy plotted on background galaxy cutout, and green colored circular region signifies the central region as defined in section \ref{3.2}. }
         \label{Fig4}
   \end{figure}

As mentioned earlier, the properties of our sample galaxies, such as the stellar mass and redshift, are taken from \cite{2020AJ....159..103K}. A few sample galaxies for which stellar masses were unavailable are removed from further analysis.
The stellar mass and redshift distributions of our final 6155 (194 post-merger and 5961 non-interacting) dwarf galaxy sample are shown in Fig. \ref{Fig3}. From the figure, it is evident that the stellar mass range of the post-merger and non-interacting samples are comparable, with the majority of the dwarf galaxies falling within the range of (M$_{*}$ $\simeq$ 10$^{8}$ -- 10$^{9.6}$ M$_{\odot}$). However, the redshift range of both samples is not similar. When the non-interacting sample covers a redshift range of 0.01 -- 0.12, there are very few post-merger dwarfs beyond the redshift of 0.06. This is expected as tidal features will be harder to discern for more distant galaxies that are fainter and poorly resolved.\\

\textbf{\subsection{FUV emission from dwarf galaxy}} \label{3.2}
The calibrated GALEX images are analysed to obtain the integrated FUV flux of our sample galaxies. The region containing the sample galaxy is extracted from the 1.2$^{\circ}$ field of the GALEX tile. This extracted image has the sample galaxy at the center and has dimensions equal to $\sim$ 2.5 $\times$ 2.5 arcmin$^2$. The chosen size of 2.5 arcmin was selected based on several criteria. First, it should be greater than 2 $\times$ R$_{25}$, as confirmed by examining a sample of massive dwarf galaxies having low redshift. 
Additionally, we visually inspected the extracted images to ensure the full coverage of the dwarf galaxies in the images. 
Furthermore, this size selection helps to remove the possible contribution of any nearby sources that may lie close to the sample dwarf galaxy within the GALEX field of view.

Our sample galaxies have an irregular shape. So we identified the largest structure corresponding to the sample galaxy using the Astrodendro \citep{2008ApJ...679.1338R} python package, which is used to compute dendrograms\footnote{\url{https://dendrograms.readthedocs.io/en/stable/}}. 
Astrodendro detects structures in an intensity map based on a specified threshold flux value and a minimum pixel count, in the form of dendrograms. \textit{Dendrogram} can be considered as a hierarchical tree diagram consisting of three types of structures that share a parent-child relationship, namely leaves, branches, and trunks. Leaves are the densest flux structures (with no further sub-structures), and a collection of connected leaves forms a branch structure (parent structure of the leaves). The branches are typically much bigger and less dense than the leaves. The dendrogram's largest and least dense structure, which hosts multiple branches and isolated leaves, is known as the trunk structure (with no parent structure).

To detect reliable structures in an observed image, astrodendro uses three parameters: min$\_$value, min$\_$delta, and min$\_$npix. The threshold flux in each pixel, above which a structure is identified, is set as min$\_$value. Setting an appropriate minimum threshold enhances the reliability of the results by filtering out noise and focusing on significant structures. We used the mean sky background plus three times of standard deviation of the sky background as the minimum threshold. These values are estimated using the GALEX FUV sky background image, corresponding to the target field, downloaded from the MAST archive.

The $min\_delta$ parameter determines if a leaf is significant enough to be considered as an independent structure. This parameter helps to discern true structures from local maxima, which could be due to random noise, by imposing a strict criterion for the inclusion of pixels to be part of a structure. We used the standard deviation of the sky background as the set $min\_delta$ value. The value of $min\_npix$ dictates the minimum number of pixels a structure must encompass in order to be recognized as an independent structure. Considering the spatial resolution and PSF of the GALEX FUV image, we have fixed this value as 10. This will ensure that the detected structures will be larger than the point spread function (PSF). The spatial resolution (4.2") of the GALEX FUV images, which corresponds to $\sim$ 1 -- 10 kpc linear sizes at the redshift of our sample galaxies, is not sufficient to detect many star-forming clumps in these galaxies in the form of leaf structures. As we aim to measure the total FUV flux associated with our target galaxies, we are only interested in the largest structure or trunk structure identified in the extracted images by the astrodendro package. The shape of the trunk structure identified in the extracted image of a sample galaxy represents the shape of the galaxy in UV. The trunk structure identified for the dwarf galaxy LEDA 1148477 is shown in Fig \ref{Fig4}.

Based on the input parameters, the astrodendro identifies structures in the extracted FUV images of our sample galaxies and provides the position, area, and flux in the unit of counts per second (CPS, as the GALEX images are in the units of CPS)
of all the identified structures. The CPS is converted to flux using the calibration relations for the GALEX FUV filter as given in \cite{2007ApJS..173..682M}. The trunk structure, corresponding to the entire galaxy, has the maximum area, and the flux of the trunk structure is taken as the total FUV flux of the target galaxy. The sky background is corrected using the corresponding FUV sky background image taken from the GALEX MAST archive. The flux corresponding to the sky is calculated by multiplying the mean sky background value by the exact area of the trunk structure, and this flux is subtracted from the total FUV flux measured from the galaxy. Then we corrected the flux for the Galactic foreground extinction (which accounts for the absorption and scattering of light by the interstellar dust within our Milky Way galaxy) using the E(B$-$V) values provided by \cite{Schlegel_1998} and applying the calibration given by \cite{Schlafly_2011}. The extinction in FUV is given as,  A$_{FUV}$ =R$_{FUV}$ $\times$ E(B$-$V) and the $R_{FUV}$ value is taken as 8.06 \citep{2011Ap&SS.335...51B}. We note that the $R_{FUV}$ values for low-mass galaxies can deviate from the assumed value of the Milky Way, to up to approximately 1.5 times higher \citep{2011Ap&SS.335...51B}. The estimated total flux, corrected for the sky background and Galactic extinction, is further used to calculate the SFR of the galaxy. This flux is not corrected for the internal extinction and hence provides the lower limit.

\textbf{\subsection{Estimation of star formation rate}}
The current/instantaneous SFR of a galaxy is proportional to the FUV emission by the young stars. Assuming that the SFR is approximately constant over the past 100 Myr, the observed FUV flux in star-forming galaxies serves as a direct indicator of current star formation activity \citep{1998ARA&A..36..189K,1998ApJ...498..106M}. To estimate the current SFR of our sample galaxies, we used the equation given by \cite{Murphy_2011} for the GALEX FUV band.
\begin{align}
        ~~~~~~~~~~~~~~~~~~SFR_{FUV}(M_{\odot} yr^{-1})=4.42 \times 10^{-44} L_{FUV} 
\end{align}

\cite{Murphy_2011} derived this relation using spectra generated by Starburst99 \citep{1999ApJS..123....3L} models, assuming instantaneous star formation, solar metallicity, a Kroupa \citep{2001MNRAS.322..231K} IMF (with a slope of $-1.3$ and $-$2.3 for stellar masses between $0.1-0.5 M_{\odot}$ and $0.5-100 M_{\odot}$ respectively) and finally convolving the spectra with the GALEX transmission curves. 
The calibration constant used in the relation can vary by up to a factor of $\sim$ 1.5 depending on stellar metallicity (at lower metallicities the UV luminosity increases; \citealt{2005A&A...443L..19B}) and we note that it can affect the absolute SFR values of our sample dwarf galaxies. But, similar mass galaxies are expected to have similar metallicity values. Therefore, our final results based on the comparison of the SFR between the post-merger and non-interacting galaxies in the same stellar mass range are not expected to be significantly affected by the choice of calibration constant. The FUV luminosity (L$_{FUV}$) of our sample galaxies is calculated using the relation,
\begin{align}
    ~~~~~~~~~~~~~~~~L_{FUV} (erg~s^{-1})=4 \pi D^{2} \times Flux(erg~s^{-1}~cm^{-2}) 
\end{align}

where D is the distance to the target dwarfs and is calculated using the spectroscopic redshift values adopted from the GAMA and SDSS spectroscopic database. 

As discussed in section 3.2, the GALEX FUV images' spatial resolution is insufficient to detect multiple star-forming regions in our target galaxies. Hence, we cannot study the effect of interactions on the spatial distribution of star formation in our sample galaxies. However, to understand the nature of star formation in the central and outer regions of these target galaxies, we estimated the SFR in the central (defined by a circle corresponding to 3" radius) and the region outside it. The central 3" radius is defined based on the spatial resolution of the GALEX image. This chosen size also allows us to make a comparison of our results (see the discussion in Section \ref{5}) with the results of \cite{2020AJ....159..103K}, who showed an increased SF activity in the central regions of these post-merger dwarfs, based on the SDSS fiber (which is of the size of 3" diameter) spectra.

   \begin{figure}
   \centering
   \includegraphics[width=9.0cm]{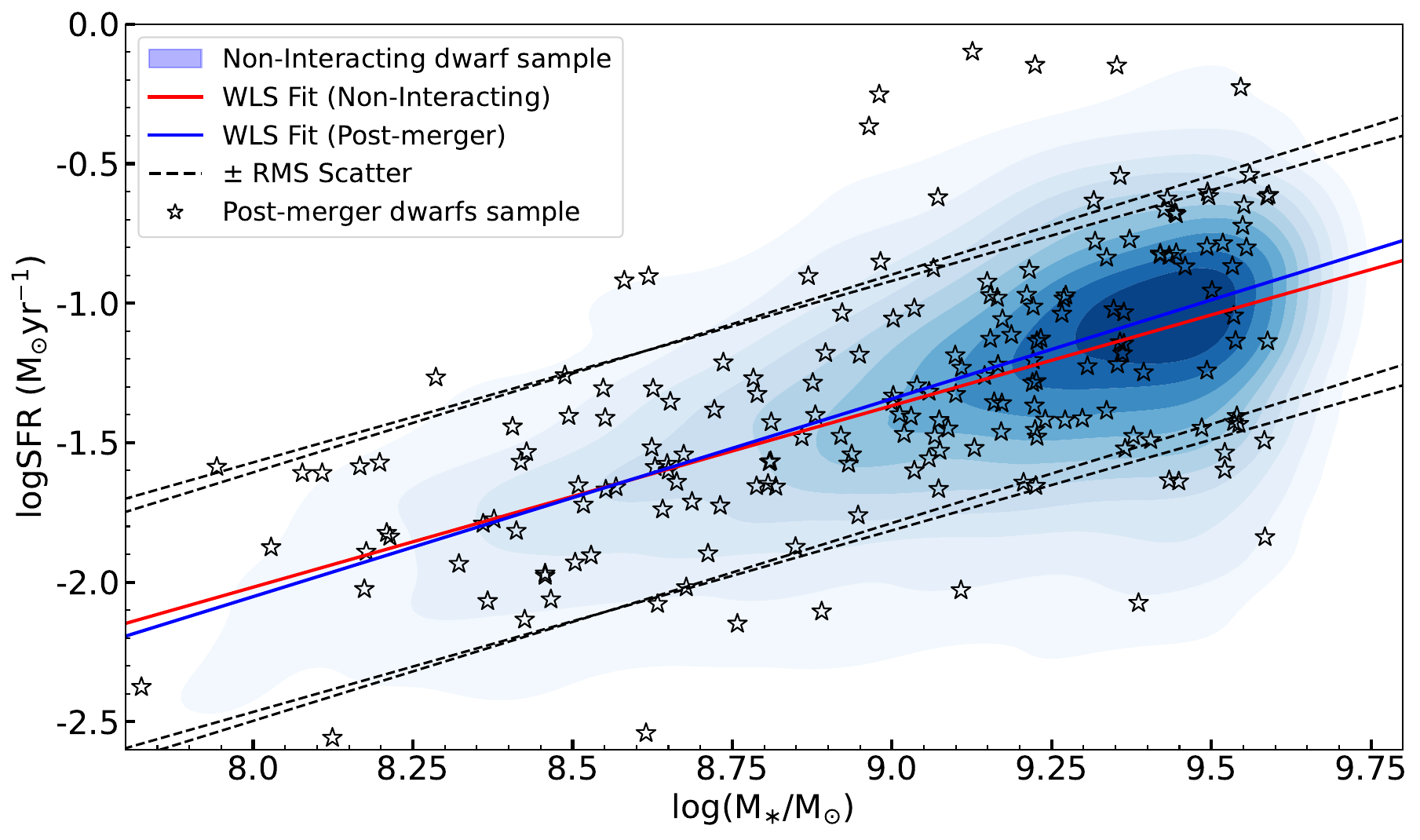}
      \caption{ SFR vs Stellar mass with density contours plotted over it for non-interacting dwarfs, while individual scatter points denote post-merger dwarfs. The figure indicates that post-merger dwarfs tend to exhibit slightly elevated SFRs compared to their non-interacting counterparts at a given stellar mass.}
         \label{Fig5}
   \end{figure}

\textbf{\subsection{Automated analysis of our sample dwarf galaxies}}
As discussed earlier, due to the large sample size (with an initial sample of 6875 galaxies) we automated all the steps starting from searching and downloading the highest exposure GALEX FUV images containing our target galaxies from the MAST archive, extracting the sample galaxy region, identifying the structure and estimating the total observed FUV flux of the sample galaxies using the astrodendro, background sky subtraction, foreground extinction correction, estimating the FUV luminosity and finally the total SFR, along with the central and outer SFR, are performed using a single python based code. The input to the code is the RA and DEC of the target galaxies. All the final estimated parameters of a total of 6155 (194 post-merger and 5961 non-interacting) dwarf galaxies are saved in the output file. The results based on these parameters are presented in the next section. \\

\section{Results} \label{4} 
We have estimated the instantaneous SFR of 6155 dwarf galaxies spanning a mass range of ($10^{7}-10^{9.6}$ $M_{\odot}$) and a redshift range of 0.01 -- 0.12. This sample of dwarfs contains 194 post-merger (which show tidal features around them in the deep optical images) and 5961 non-interacting dwarf galaxies. To investigate the impact of interactions on the star formation properties of dwarf galaxies, we compare the estimated SFR of the post-merger and non-interacting samples of dwarf galaxies in similar mass and redshift bins. We also compare the SFR in the central region (corresponding to a 3" radius circular region) and the outer region of these galaxies. 
 \begin{figure*}[ht!]
   \centering
   \includegraphics[width=0.49\textwidth, height=0.35\textwidth]{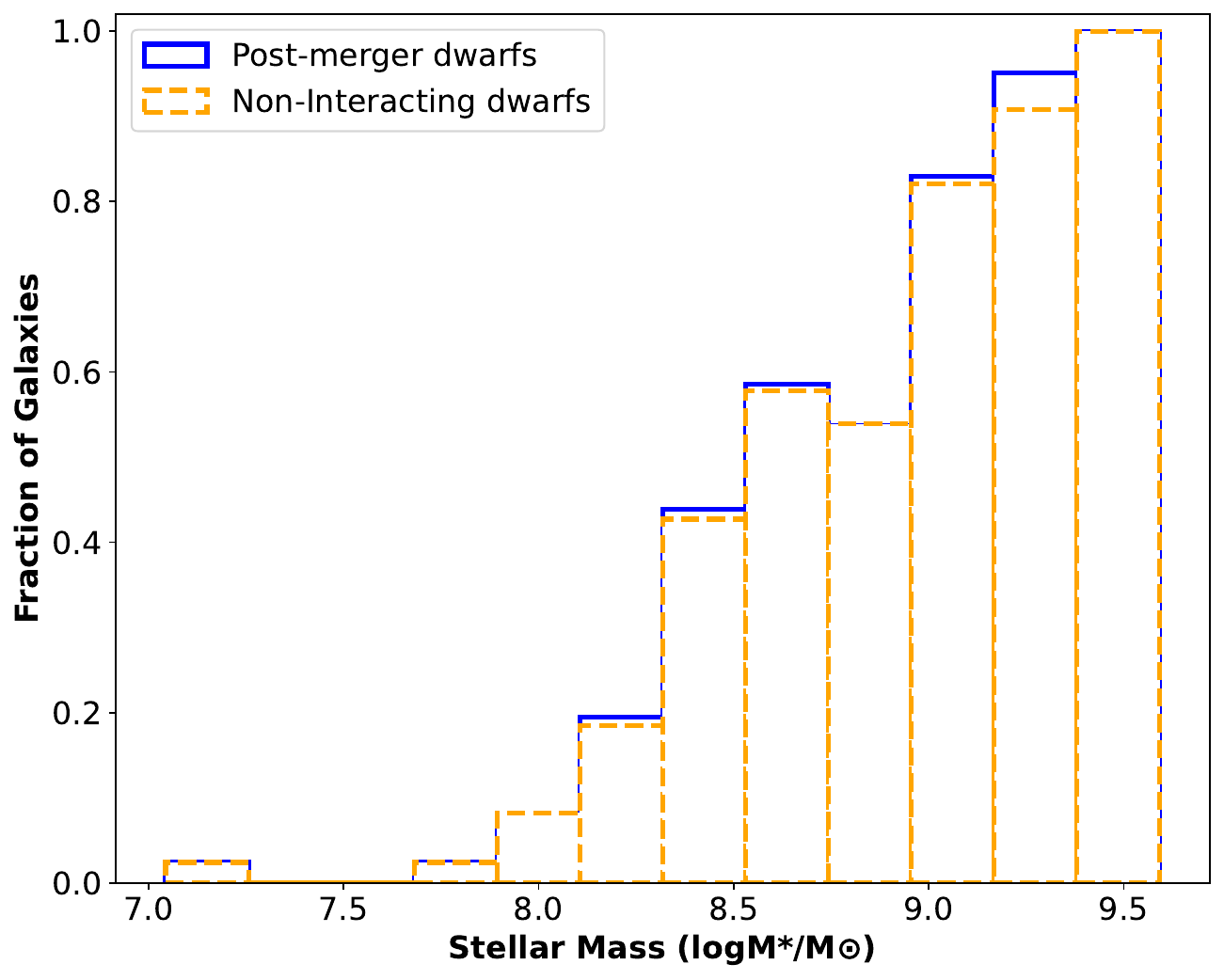}
    \includegraphics[width=0.50\textwidth,height=0.35\textwidth]{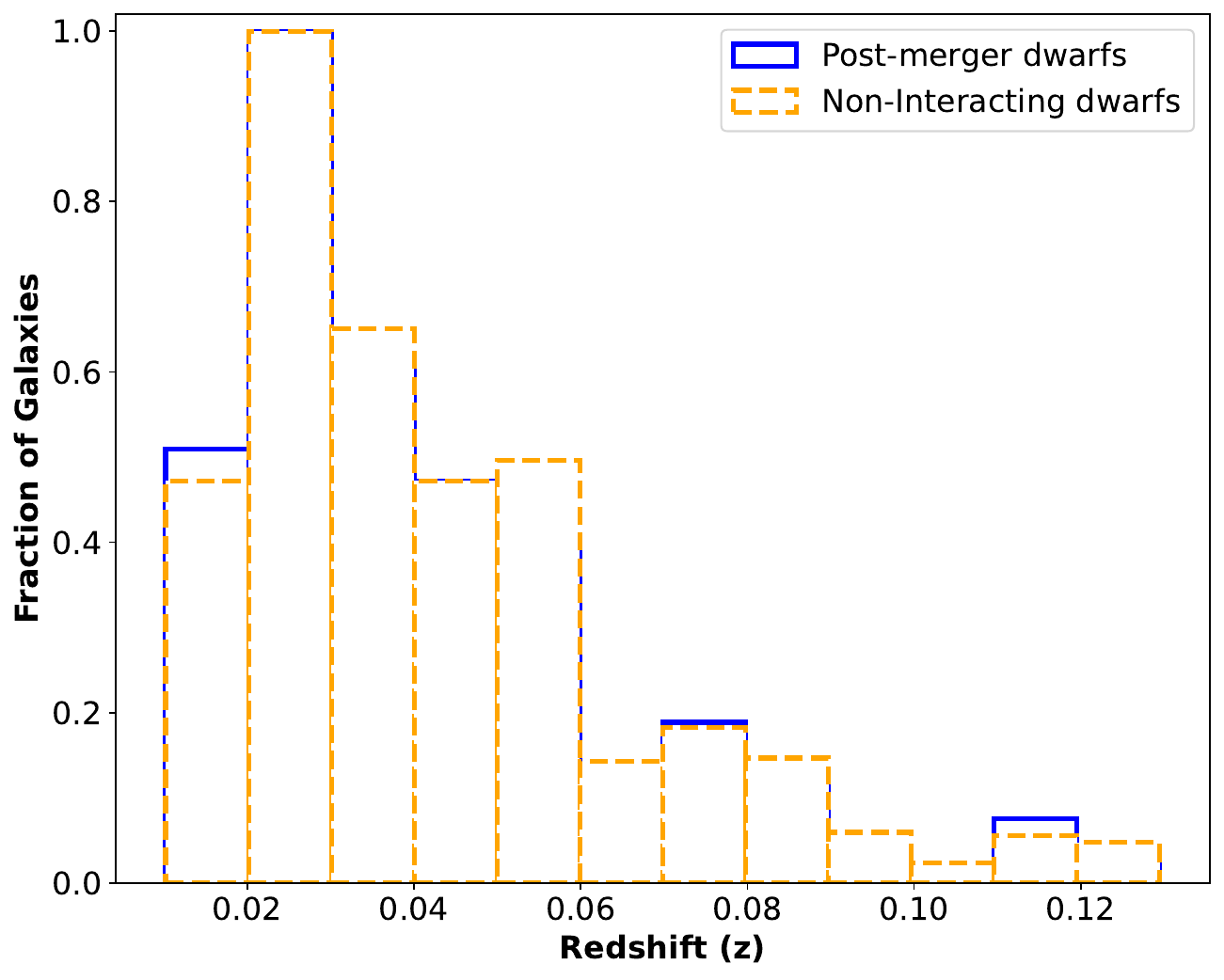}
  \caption {Histogram showing the distribution of post-merger and non-interacting control dwarfs corresponding to stellar mass (left) and redshift (right). Post-merger galaxies are shown with blue solid lines, while non-interacting dwarfs are shown with orange dashed lines.}
         \label{Fig6}
   \end{figure*}
The specific Star Formation Rate (sSFR, which is SFR/stellar mass of the galaxy) of the post-merger and non-interacting samples is also compared, as a function of redshift. 

\textbf{\subsection{Star Formation Rate}} \label{4.1}
The star-forming sequence (SFR vs. stellar mass) of galaxies is a key observational tool for understanding the assembly process and evolution of galaxies. Fig. \ref {Fig5} shows the star formation sequence of our sample galaxies. The background shows the log(SFR) versus log(stellar mass) for the non-interacting sample, as a density plot. The post-merger galaxies are overplotted as star symbols. As expected, the SFR of both the post-merger and non-interacting samples increases as a function of stellar mass. We performed a weighted least-squares (WLS) linear fit. The fits were applied separately to the post-merger and non-interacting dwarf galaxy samples.
The equations for the best-fit lines  for the post-merger (eqn. \ref{0.1}) and non-interacting (eqn. \ref{0.2}) samples are given below:\\
\begin{align}
     ~~~log(SFR) = (0.709 \pm 0.055)~log(M_{*}) + (-7.724 \pm 0.500) \label{0.1} \\   
     ~~~log(SFR) = (0.651 \pm 0.012)~log(M_{*}) + (-7.222 \pm 0.114) \label{0.2} 
\end{align}\\
Though the fit indicates a slightly steeper slope and lower intercept for post-merger galaxies, both these values are similar (within errors) to those obtained for the non-interacting sample. The best-fit line corresponding to the non-interacting galaxies and $\pm$ rms (0.447 dex) lines are shown in Fig. \ref{Fig5} as red and dotted green lines respectively. Similarly, the best-fit and the rms (0.445 dex) lines for the post-merger galaxies are shown as blue and dotted black lines. 

The SFR of galaxies increases not only as a function of their stellar mass but also as a function of redshift. Hence, to make a meaningful comparison between the SFR of the post-merger and non-interacting dwarfs, we constructed a control sample simultaneously matched in stellar mass and redshift. For each post-merger dwarf, we select control counterparts from the non-interacting population within the mass–redshift parameter space, ensuring no replacement and iteratively identifying the next-best matches. To ensure statistical consistency, we ran a Kolmogorov–Smirnov (KS) test between the post-merger and control samples, setting a criterion that the KS probability is greater than a 30\% level (e.g., \citealt{2008AJ....135.1877E,2011MNRAS.418.2043E}). We aimed to find a match for a maximum of 5 control non-interacting counterparts for each post-merger dwarf. Eight post-merger dwarfs did not yield the required number of matched control dwarfs under our selection criterion. We relaxed the KS probability threshold to 25\% for those cases to obtain the corresponding matched control dwarfs.
Eventually, we compare a sample of 194 post-merger dwarf galaxies with a unique control sample of 970 non-interacting dwarfs, selected based on a KS probability threshold. 
The corresponding distributions of the post-merger and control samples are presented in Fig. \ref {Fig6}.

   \begin{figure*}
   \centering
   \includegraphics[width=0.49\textwidth, height=0.35\textwidth]{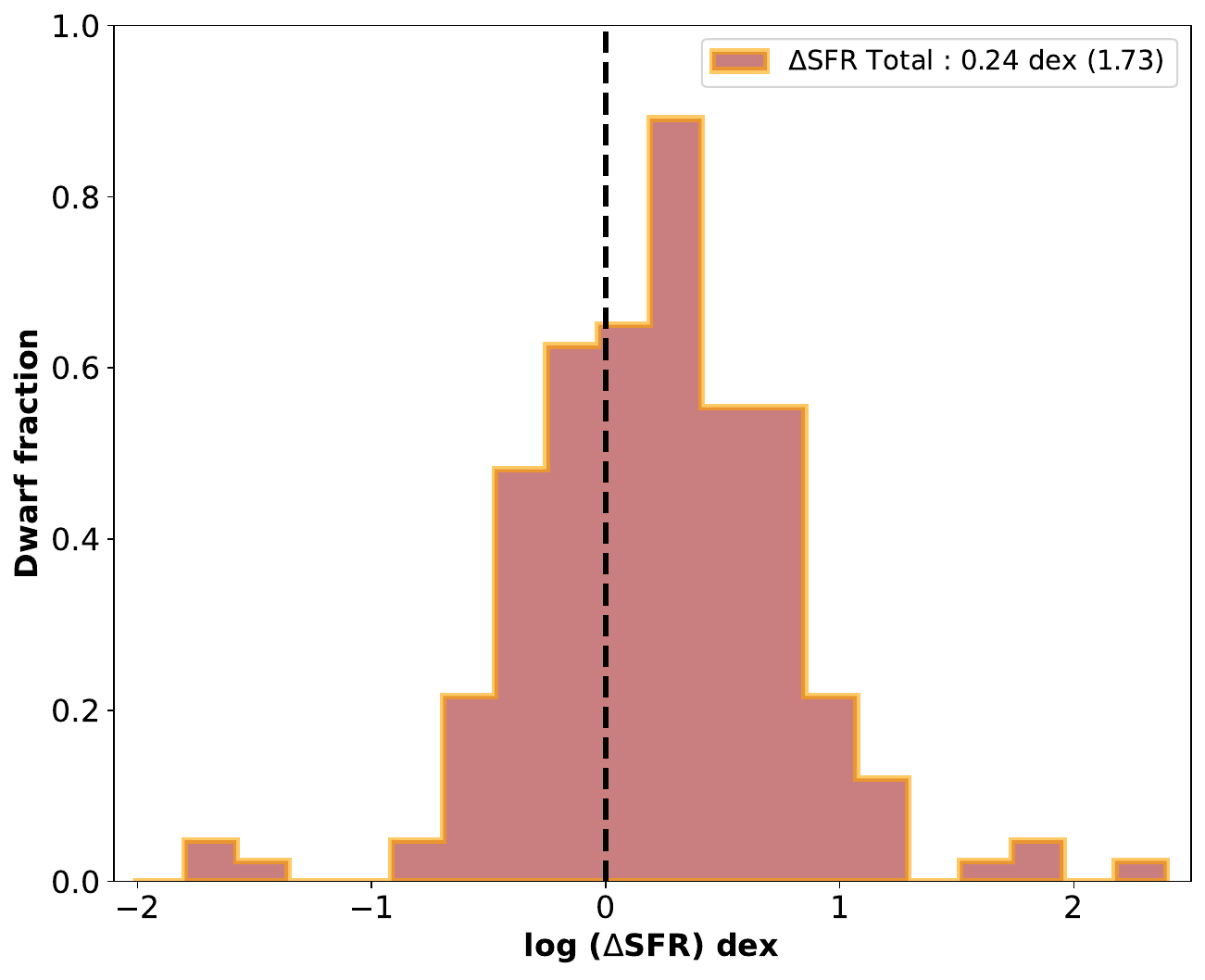}
    \includegraphics[width=0.50\textwidth,height=0.345\textwidth]{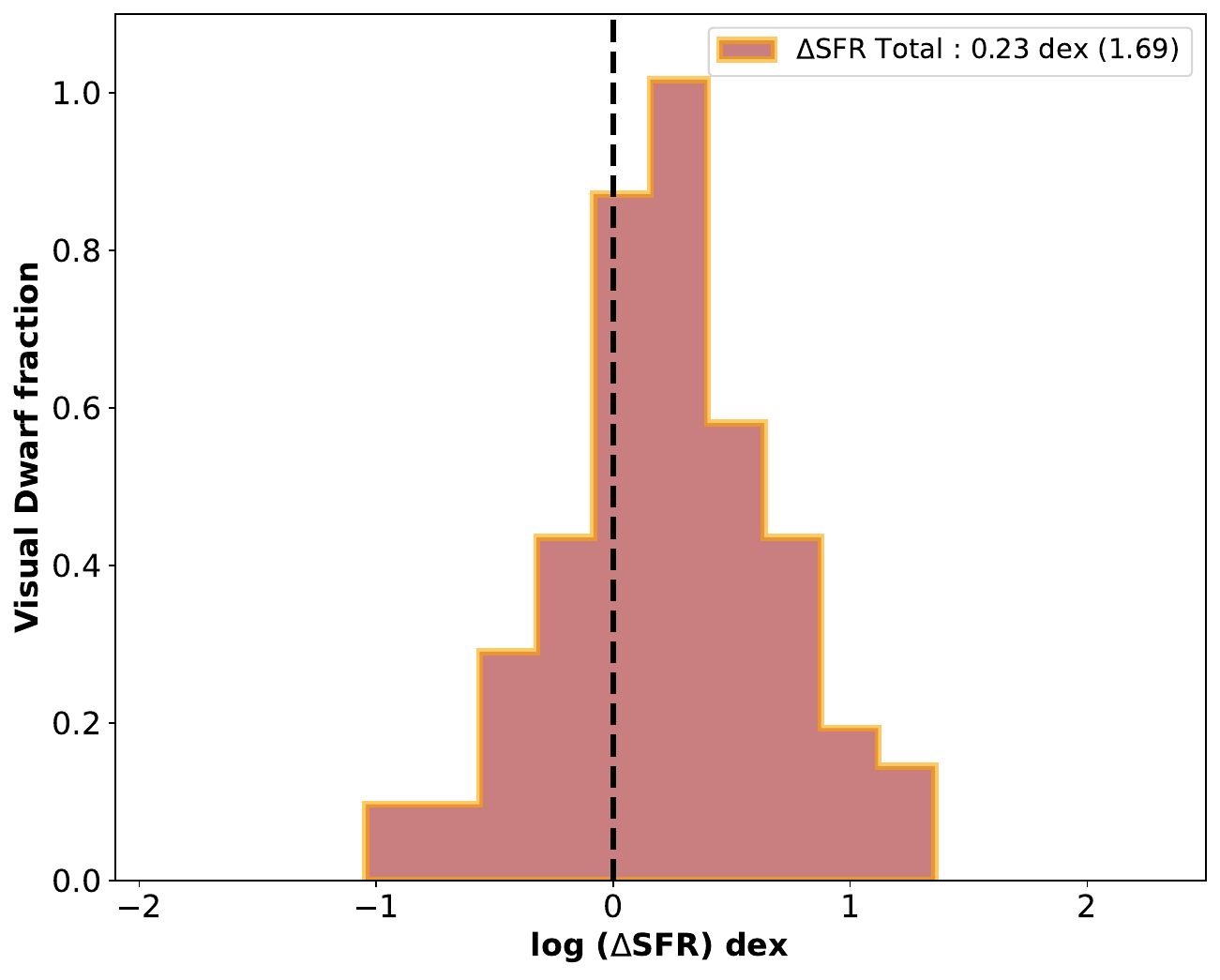}
    \includegraphics[width=0.49\textwidth, height=0.35\textwidth]{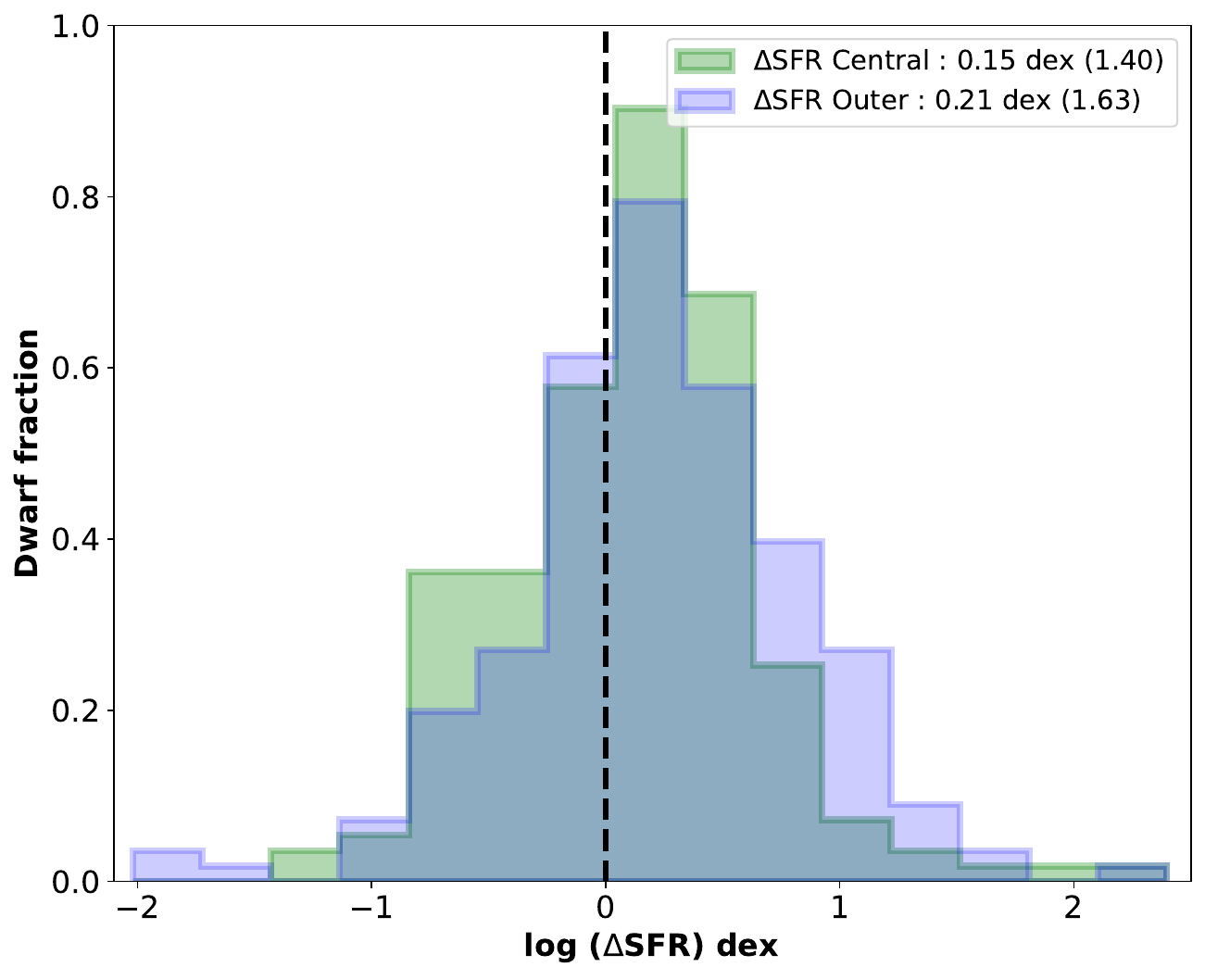}
    \includegraphics[width=0.50\textwidth, height=0.345\textwidth]{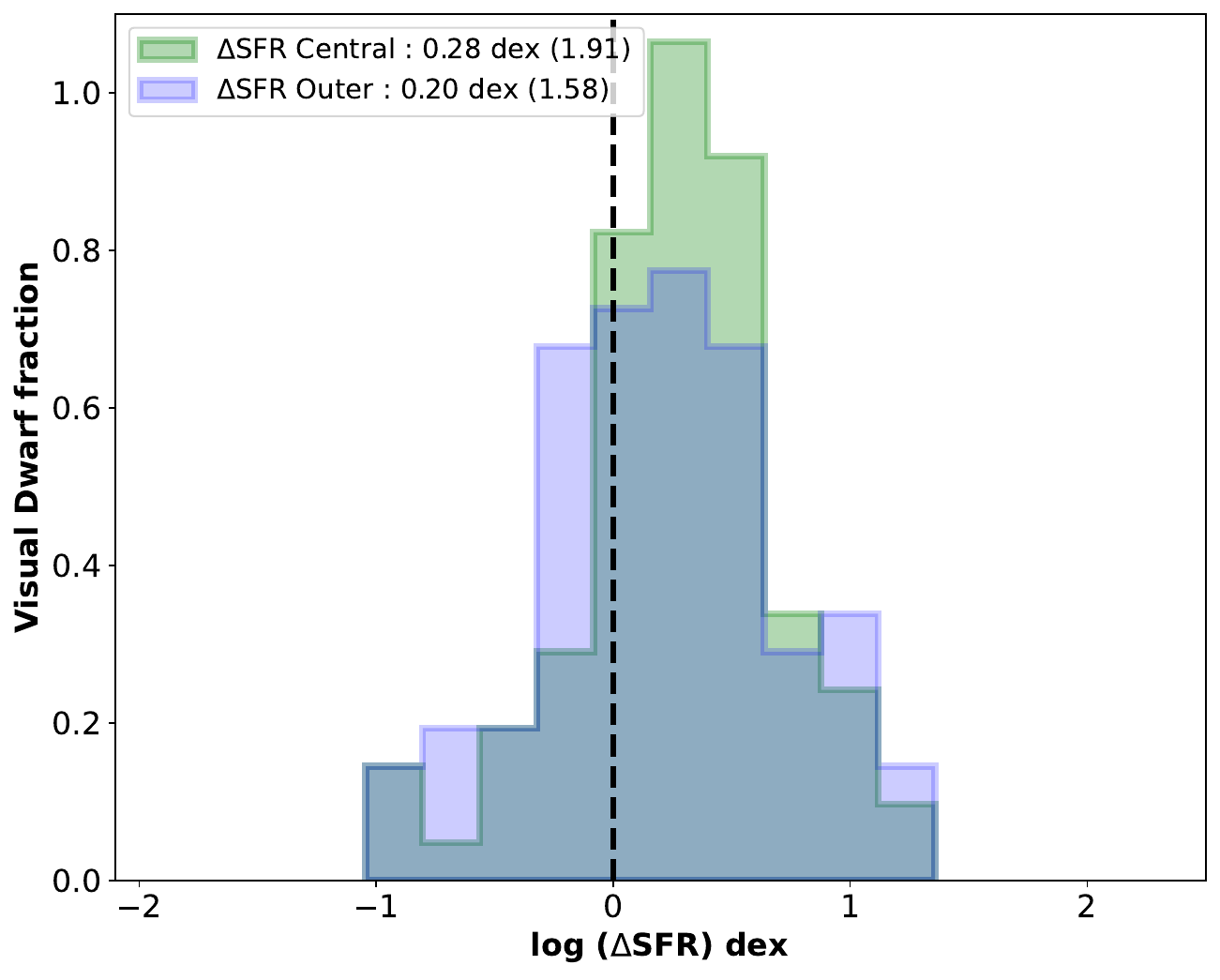}
      \caption{The plots comparing the difference in the SFR  (log($\Delta$SFR) =  logSFR$_{post-merger}$ $-$  logSFR$_{non-interacting}$) of the post-merger with respect to non-interacting control dwarfs. Left panel: This figure shows the histogram distribution of SFR offset for the entire galaxy, central, and outer regions. Right panel: This figure shows the histogram distribution of SFR offset for the visually identified dwarf galaxy sample, central and outer regions.} 
         \label{Fig7}
   \end{figure*}

To quantify the effect of mergers on SFR,  we calculated the logarithmic ratio between the SFR of each post-merger system and the median SFR of its matched control sample. On a logarithmic scale, this ratio is the offset in log(SFR) between a post-merger galaxy and the median of its corresponding control sample. The upper-left panel of Fig. \ref{Fig7} shows the distribution of this offset for our 194 sample of post-merger galaxies. Though the ratio/offset has a range ($-$2 to $+$2 dex, about 100 times suppression/enhancement), indicating both enhancement and suppression of star formation in these recent merger galaxies, around 67\% of the sample (130 galaxies) shows an offset greater than 0, indicating that most of the sample galaxies show an enhancement in SFR. The median offset (enhancement) of the sample is 0.24 dex (1.73), indicating a $\sim$ 70\% increase in the SFR of recent merger galaxies compared to their non-interacting counterparts. Out of 130 galaxies that show SFR enhancement, 66\%, 30\%, and 13\% show 2, 5, and 10 times enhancement in SFR, respectively. We further checked these factors only for those 86 galaxies that show tidal features in visual classification. As they are identified in visual classification, they represent the galaxies with strong features from the recent major merger event. The upper-right panel of Fig. \ref{Fig7} shows the SFR offset distribution of 86 post-merger sample galaxies. The range of offset ($-$1 to $+$1 dex) is narrower than that observed in the total sample. However, the median enhancement value remains comparable, at 0.23 dex (a factor of 1.69). Among the visually identified dwarf galaxies, approximately 71\% exhibit enhanced star formation. Specifically, within a subsample of 61 dwarfs, 57\%, 24\%, and 5\% show enhancements in SFR by factors of 2, 5, and 10, respectively.

We find that there is an overall enhancement of $\sim$ 1.7 times in the SFR of post-merger dwarf galaxies, identified either through visual classification and/or an automated detection algorithm, relative to their non-interacting control counterparts. The sample of post-merger galaxies ($\sim$ 33\%) that show suppression (offset $<$ 0) of star formation might be either at a late stage of merger, where quenching of star formation can happen as seen in massive galaxies (\citealt{2024OJAp....7E.121E,2025MNRAS.538L..31F}), and/or affected by secular effects. However, quenching of star formation in the post-merger regime of dwarf galaxies is not well understood.

\textbf{\subsection{Spatial Distribution of Star formation}} \label{4.2}
As discussed earlier, to understand the effect of interactions on the spatial distribution of star formation, we compared the SFR in the central (within an aperture with a 3" radius) and outer (outside of the 3" radius aperture) regions of the post-merger sample and their non-interacting control counterparts. The lower panels of Fig. \ref{Fig7} show the distribution of SFR offset in the central and outer regions of the post-merger sample. The lower-left and lower-right show the distributions for the total (194) and only visually identified (86) sample of post-merger galaxies. The plots suggest that there is a similar SFR enhancement in the central and outer regions of the post-merger dwarfs. We further quantify the level of SFR enhancement in the post-merger dwarf galaxy population, finding that the SFR is elevated by factors of 1.40 and 1.63 in the central and outer regions, respectively. For the visually identified sub-sample, the central and outer enhancement factors are 1.91 and 1.58, respectively. These results indicate that the enhancement is not confined to the galactic center but is instead distributed broadly throughout the galaxy, suggesting a global response to star formation resulting from interactions.

\indent Moreover, we also investigated the potential dependence of SFR enhancement within different stellar mass bins of $ 10^{8-9} M_{\odot}$ and $10^{9-9.6} M_{\odot} $ in the central, outer, and overall regions of post-merger dwarf galaxies relative to non-interacting control sample. The corresponding plots illustrating the SFR enhancement across two stellar mass bins are presented in the appendix in Fig. \ref{Fig9}. However, no significant variation in the enhancement factors (for the total, central and outer) is observed between the two stellar mass bins. These result suggests that interactions among dwarf galaxies can overall enhance the star formation activity, while the enhancement in SFR is largely independent of their stellar mass. As the number of galaxies in the mass range of $ 10^{7-8} M_{\odot}$ is less, we did not perform a similar analysis for that mass bin.

   \begin{figure*}
   \centering
   \includegraphics[width=0.49\textwidth, height=0.35\textwidth]{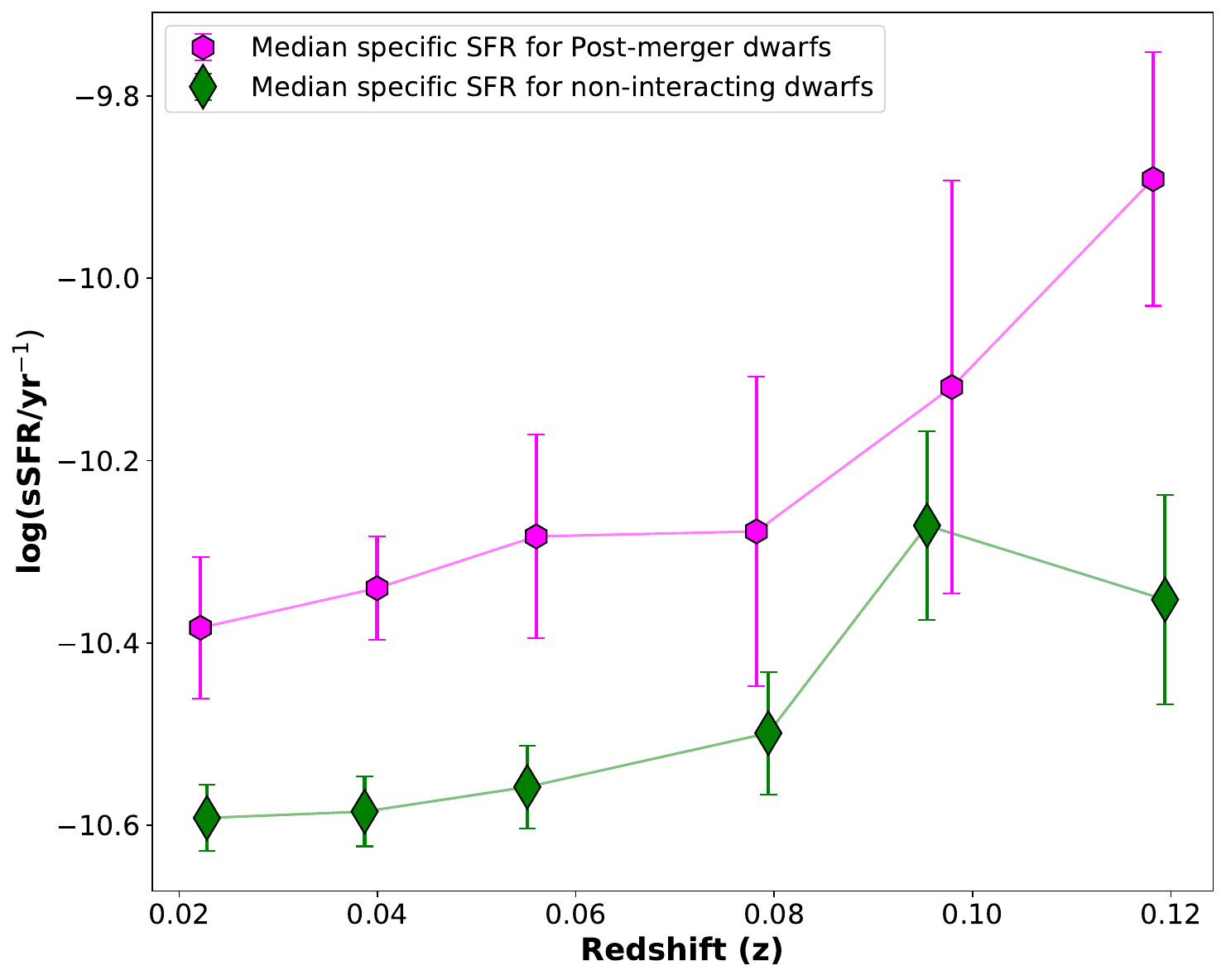}
    \includegraphics[width=0.50\textwidth,height=0.35\textwidth]{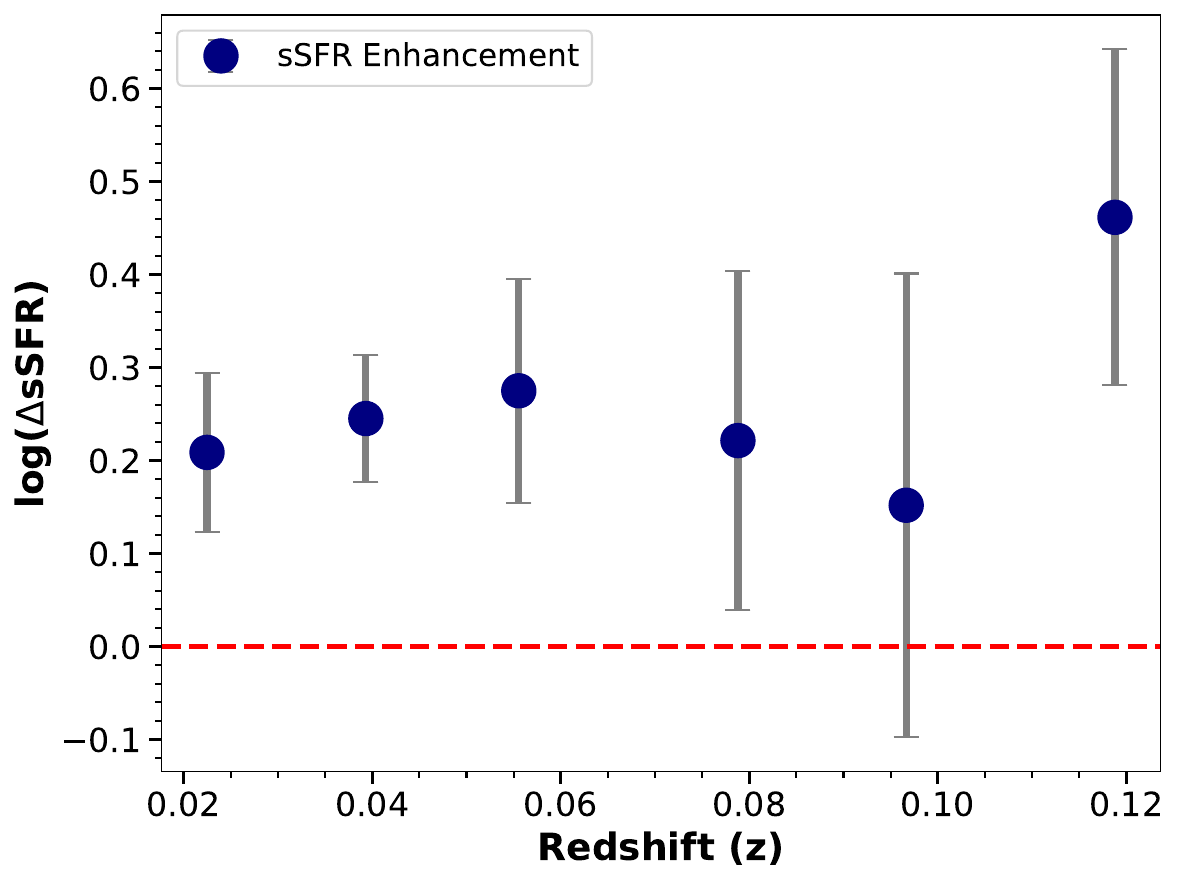}
     \caption{Left panel: sSFR vs redshift plot which shows higher median sSFR for the post-merger dwarfs, also a mild increasing trend beyond z $\sim$ 0.08 for post-mergers. Right panel: This plot shows the offset in sSFR, i.e., (log($\Delta$sSFR) =  logsSFR$_{post-merger}$ $-$ logsSFR$_{non-interacting}$) between the post-merger and the corresponding non-interacting control dwarfs, as a function of redshift. The red dotted line denotes log($\Delta$sSFR) = 0. This plot demonstrates an enhancement in sSFR for post-merger dwarfs relative to matched controls at all redshifts. }
         \label{Fig8}
   \end{figure*}

\textbf{\subsection{Specific Star Formation Rate}} \label{4.3}
In this subsection, we analyze the variation of specific star formation rate (sSFR, which normalises SFR by stellar mass) of our sample of dwarf galaxies as a function of redshift. The left panel of Fig. \ref{Fig8} shows the variation of sSFR as a function of redshift for both post-merger dwarfs and their non-interacting control counterparts. It demonstrates that post-merger galaxies consistently exhibit higher sSFR values than non-interacting dwarfs across all redshift bins. The right panel of Fig. \ref{Fig8} illustrates the offset between the sSFR (log($\Delta$sSFR) =  log(sSFR$_{post-merger}$) $-$ log(sSFR$_{non-interacting}$)) as a function of redshift. Notably, this offset remains above the line log($\Delta$sSFR) = 0, indicating a systematic elevation in post-merger systems relative to matched controls, throughout the redshift range. However, given the relatively narrow redshift interval probed, it is unclear whether this enhancement is a global trend. Additionally, the left panel shows that sSFR remains nearly constant up to z $\approx$ 0.08 for both samples, with a mild increase observed at higher redshifts for post-merger galaxies. Whereas, the non-interacting dwarfs display more fluctuations beyond z > 0.08. These observed features might be due to the smaller sample size and the incompleteness of the sample at higher redshifts, as discussed in Sections \ref{3.1} and \ref{5}.  \\

\section{Discussion} \label{5}
The process of star formation within galaxies is crucial for transforming gas into stars, significantly influencing the evolutionary stages of galaxies. However, the effect of dwarf-dwarf interactions on their star formation properties remains poorly understood. The significance of interactions and star formation in galaxy growth changes depending on stellar mass and redshift. Galaxies stellar mass is strongly correlated with their kinematical and morphological properties \citep{2005ApJ...625..621B,2010A&A...522A..18B}, and linked to the characteristics of their stellar populations and interstellar medium (ISM), such as through the mass–metallicity relation \citep{2004ApJ...613..898T}. 
Therefore, accurately estimating SFR and relating it to stellar mass is essential to study a wide range of galaxies, particularly to investigate the evolutionary trajectories of these systems. 
Star-forming galaxies are generally expected to follow a tight relationship with M$_{*} \propto$ SFR, which evolves gradually with redshift. This has been explored both observationally \citep{2007ApJ...670..156D,2007A&A...468...33E} and theoretically \citep{2006MNRAS.370..273D,2007MNRAS.376.1861F} and is known as the star-forming main sequence (SFMS).  

By studying the evolution of the SFMS over cosmic time, we can trace the pathways through which galaxies have evolved. Galaxies that lie above the main sequence are typically known to be undergoing interactions or mergers, which perturb the gas and trigger an increase in their star formation rate. Our sample of post-merger dwarfs is distributed uniformly along the SFMS, exhibiting SFRs comparable to those of their non-interacting counterparts across a similar stellar mass range. This indicates that our sample is not dominated by starburst or quenched systems as a result of the merger.
Although dwarf mergers are found on an elevated main sequence, exhibiting higher SFRs for a given stellar mass \citep{2017AAS...22912302S} similar to their more massive counterparts.
Low-mass galaxies typically experience more quiescent phases than their more massive counterparts, up to an average of 2.8 times throughout their lifespan. However, these quiescent phases are typically shorter, lasting on average only 1.49 billion years \citep{2023MNRAS.522L..11H}. The findings of \citealp{2017ApJ...851...22M} suggest LSB dwarfs exhibit a steep slope of (1.04 $\pm$ 0.06) while \cite{2018ApJ...864L..42K} report a slope near unity for dwarfs in the redshift range 0.5 < z <3. \cite{2024A&A...681A...8S} found a slope of 0.62 and an intercept of $-$ 7.4 for the SFMS of isolated dwarf galaxies, (22 interacting and 36 non-interacting within a mass range of 10$^6$ -- 10$^9$ M$_{\odot}$), in the Lynx Cancer Void. They also found that both the interacting and non-interacting samples share similar SFMS. Our analysis yields similar values for the slopes and intercepts. \cite{2024MNRAS.530.2199A} used the mass and SFR of star-forming clumps in 16 star-forming dwarf galaxies (stellar mass $<$ 10$^9$ M$_{\odot}$), and reported a slope of 0.899 $\pm$ 0.087 and an intercept of -8.59 $\pm$ 0.75 for star-forming dwarf galaxies. The mass of the star-forming clumps ranges from 10$^3$ - 10$^6$ M$_{\odot}$ and hence this relation traces the lower mass regime. 

\citet{2015ApJ...801L..29R} focused on the massive star-forming galaxies within the spectroscopic redshift range 0.02 < z < 0.085, and reported a slope of 0.76 and an intercept of –7.64 values that closely align with those derived from our sample. Consistent results have also been reported by other studies, such as \cite{2012ApJ...754L..29W,2016ApJ...821L..26C}, which, although primarily targeting more massive systems, cover a broader stellar mass range and span redshift ranges comparable to those of our dataset.

Our study found that star formation triggered as a result of coalescence approximately doubles the rate of stellar mass growth, relative to a sample of matched control non-interacting counterparts. 
A study by \cite{2015MNRAS.454.1742K} took a sample of nearly 1500 galaxies having M$_{*} \sim 10^{8-11} M_{\odot}$ and lying within distance of $\sim$ 45 Mpc. They found out that both SFR and sSFR (discussed in \ref{4.2}) are enhanced in interacting galaxies. Although their sample covers a larger range of stellar mass, the sample consists of dwarfs within M$_{*}$ of $10^{8-10} M_{\odot}$. The enhancement was moderate and increased with the degree of interaction, reaching up to maximum enhancement by a factor of 1.9 times for the highest degree of interaction (mergers). \citet{2019A&A...631A..51P} demonstrated that galaxy mergers do have an impact on the SFR of the interacting systems, although the effect is relatively modest, with a typical enhancement factor of approximately 1.2. Studies based on the observations of massive galaxies \citep{2013MNRAS.435.3627E,10.1093/mnras/stac1500,2025MNRAS.538L..31F} and those based on simulations by \citet{2020MNRAS.493.3716H} report that star formation in post-merger galaxies is found to be elevated. These studies find that post-merger systems exhibit SFRs approximately twice as high as those of control galaxies matched in stellar mass and redshift.
Our results based on post-merger dwarf galaxies are in agreement with the previous studies. 

The stage and nature of galaxy interactions, whether major or minor mergers or non-merger interactions, can influence the SFR. Major galaxy mergers, in particular, are known to trigger intense nuclear starbursts \citep{1994ApJ...431L...9M,2000MNRAS.312..859S,2006ApJ...650..791C}. However, \cite{2007A&A...468...61D} argue that mergers do not always lead to starbursts, and galaxy interactions alone are not sufficient for efficiently converting large amounts of gas into new stars. Instantaneous SFRs do not provide the complete picture of star formation throughout the merger sequence.
The peak of the SFR typically occurs in the later stages of a galaxy merger. 
In contrast, during the initial peri-center passage, the increase in SFR is relatively modest \citep{2007A&A...468...61D}, reaching only 4 -- 5 times the rate observed in isolated ones. The findings of \cite{10.1093/mnras/stv1500} also align with the prevailing view of current theoretical literature, suggesting that major mergers play a comparatively weaker role than minor mergers in shaping galaxy properties such as mass, size, and population gradients at early cosmic epochs \citep{Oser_2012,2014ApJ...790L..32R,2016MNRAS.461.1760H}. In the dwarf regime, mergers lead to a moderate increase in star formation by a factor of 3 –- 4 times at z = 1, while non-merger interactions result in a smaller enhancement of approximately 2 times \citep{2021MNRAS.500.4937M}. Our post-merger galaxies are suggested to be the result of a recent major merger event \citep{2020AJ....159..103K}.

Though the median SFR offset observed in our sample of post-merger dwarf galaxies, compared to their non-interacting control sample, is broadly consistent with that of an overall enhancement, the histogram reveals a wide distribution in enhancement values, ranging from approximately $-$2 to $+$2 dex. This indicates that our post-merger sample likely comprises a mix of quenched and star-bursting dwarfs. 
Observational studies have shown that the duration of interaction-induced star formation can span several hundred million years \citep{2000ApJ...530..660B, 2009ApJ...698.1437K, Woods_2010}, while numerical simulations suggest that the most intense starburst activity typically occurs near the time of coalescence \citep{1996ApJ...464..641M, 2006ApJ...650..791C, 2007A&A...468...61D, 2010A&A...518A..56M}. Since the selection of our post-merger sample is based on the detection of tidal features, potentially originating from recent or post-merger dwarf galaxies, these features can persist for up to $\sim$ 1 Gyr; our sample may consist of galaxies in different stages of the post-merger regime.  
\cite{2025MNRAS.538L..31F} showed that SFR in massive galaxy mergers can be elevated by a factor of two relative to matched control samples. Using the Multi-Model Merger Identifier (MUMMI, \citealt{2024MNRAS.533.2547F}) neural network ensemble, they were able to predict the time since coalescence for their sample galaxy. They found that the SFR enhancement persists for at least 500 Myr following coalescence, after which it declines rapidly over the subsequent 500 Myr, eventually returning to baseline levels and transitioning into a quenching phase. Among the 130 galaxies in our sample, there is an enhancement in the SFR.  Out of the total 194 galaxies, we find that 44\%, 20\%, and 9\% exhibit SFR enhancement by factors of at least 2, 5, and 10 times, respectively, which may be the result of equal mass mergers. These starburst fractions are comparable with the findings of \cite{2008A&A...492...31D}, who reported that strong starbursts with SFR enhancement factors of at least 5 are relatively rare, occurring in approximately 15\% of major galaxy interactions and mergers. 33\% of the galaxies in our post-merger dwarf sample are quenched. These systems may represent a more advanced stage of the post-merger evolutionary sequence, wherein quenching processes such as the depletion of cold gas are likely driven by mechanisms like tidal stripping during interactions. Since dwarfs have a shallower gravitational potential compared to massive galaxies, cold gas (which is the fuel for star formation) may be removed from either or both galaxies during close encounters might contribute to quenching. This gas removal can significantly suppress star formation activity. Additionally, AGN feedback may also contribute to the quenching of star formation in these dwarfs \citep{2023ApJ...957...16S}, as observed in more massive systems. However, we do not have estimates of time after the merger of these dwarf systems. Another possibility is that the morphological features seen in these quenched dwarfs might be associated with minor mergers or the coalescence of low-mass satellite galaxies. As these systems are isolated (no massive galaxies within a volume of 1 Mpc$^3$), environmental effects such as ram-pressure stripping or strangulation/starvation \citep{2014MNRAS.442.1396W} are unlikely to contribute to the suppression of star formation in these systems.

In this study, we explored the effect of dwarf mergers on the spatial distribution of star formation. We compared the star formation in the inner (3" radius circular region centred on the galaxy centre) and outer (outside of 3" radius circular region) of our post-merger sample with the control sample. We observed FUV emission across the whole galaxy. Among the 67\% of the sample, which shows enhancement in their total SFR compared to their control sample, it is also found that the SFR enhancement in the inner and outer regions is found to be similar. This suggests that the merger-triggered star formation in dwarf galaxies is not just confined to the central regions, but it also triggers star formation across the galaxy. 
This is in contrast to what is observed in massive galaxies, where the bulk of the interaction-triggered star formation is found to be centrally concentrated. Both in the pre-coalescence as well as in the post-merger phase of massive galaxies, the SFR enhancement in the outer regions is found to be significantly less compared to the central enhancement \citep{2013MNRAS.435.3627E}. Our result is consistent with the study by \cite{2017ApJ...846...74P}, who observed widespread and clumpy star formation in the dwarf galaxy pair dm1647+21, and suggested that star formation in dwarf-dwarf mergers is driven by large-scale ISM compression rather than the gas condensation within the central regions. 

Our result also appears to be consistent with a recent study \citep{2024MNRAS.533.3771L} of 211 dwarfs (with a stellar mass range of 10$^{8}$M$_{\odot}$ < M$_{*}$ < 10$^{9.5}$M$_{\odot}$ and a redshift range of z < 0.08), which found that interacting sample are bluer compared to their non-interacting counterparts at all radii. 
However, they found that the most pronounced blueward offset is in the central regions ($<$ 0.5 $\times$ effective radius, R$_{e}$), and hence, the star formation enhancement is largest in the central regions. \citet{2020ApJ...894...57S} 
found an average SFR enhancement of 1.75 $\pm$ 0.95 in the central regions of low-mass galaxies
(8 $\leq$ log$M_{*}/M_{\odot}$ $\leq$ 10). 
Additionally, they observed average SFR enhancement of f$_{\Delta SF}$ = 1.30 $\pm$ 0.61, 1.23 $\pm$ 0.60 and 0.71 $\pm$ 0.34 in the inner (0–0.5 R$_{e}$), middle (0.5–1.0 R$_{e}$) and outer (1.0–1.5 R$_{e}$) radial bins respectively, suggesting a radially decreasing trend of SFR enhancement in these galaxies. Our sample does not show a significant difference in the SFR enhancement in the inner and outer regions. 
However, due to the resolution limitations ($\sim$1–10 kpc) at the redshifts of our sample, we are unable to resolve our sample dwarf galaxies into distinct radial bins or to localize specific star-forming regions within them. Future high spatial resolution UV observations and/or optical IFU observations of our sample galaxies will help to better understand the spatial distribution of star formation. We note that our sample of interacting galaxies is representative of the post-merger sample, and most of the previous studies of dwarf-dwarf interactions are on pairs (eg, \citealt{2020ApJ...894...57S}). So, the observed differences in the spatial variation of SFR enhancement could be due to the differences in the stage of interaction of the sample dwarf galaxies.

At every stellar mass, the growth rate due to star formation increases sharply with higher redshift. sSFRs decrease with increasing stellar mass, a trend that holds not only at z=0 but also at higher redshifts \citep{2009PhDT.......233G}. Based on the findings of \cite{2010MNRAS.402.1599S}, it can be concluded that star formation occurs with slightly higher efficiency at high redshifts. Several feedback mechanisms could be responsible, such as feedback from supernovae (SNe), which are effective in expelling gas from the shallow potential wells of star-forming dwarfs. This process plays a crucial role in regulating star formation, particularly up to redshift z = 6. The variation of the sSFR with respect to redshift suggests a direct correlation between the two. \cite{2017A&A...606A.115H} investigated the mass-metallicity relation and its evolution with redshift for local dwarf galaxies, finding no significant correlation between the mass-metallicity relation and redshift. However, when SFR was included as an additional parameter, they observed a clear dependence: the coefficient for stellar mass decreased with increasing redshift. In contrast, the coefficient for the SFR remained nearly constant. In our study, we found that the sSFR at all redshift bins is higher for post-merger galaxies compared to their control sample of non-interacting galaxies. We also found a mild indication for an evolution (above a redshift of 0.08) of sSFR with redshift for post-merger galaxies, whereas the variation was comparatively greater for the non-interacting systems. We note that these observed variations/trends might be due to the smaller sample size and the incompleteness of the sample at redshifts above 0.08. At higher redshifts, the detection of faint tidal features becomes more challenging, resulting in fewer identified post-merger dwarfs in this regime. The post-merger systems that are detected at higher redshifts are likely the outcome of advanced major mergers, which tend to display more prominent features. Therefore, these detected post-mergers are expected to have elevated sSFRs relative to their matched non-interacting counterparts. This selection effect may contribute to the redshift evolution observed in Fig. \ref{Fig8}.
However, there are studies that suggest the evolution of sSFR as a function of redshift at lower redshifts. \cite{2020MNRAS.494.4969P} examined sSFR enhancement in massive galaxies over the redshift range z $<$ 1. Their study found no significant evolution in the mean sSFR enhancement with redshift within this range. However, when dividing their sample into five equally spaced redshift bins, they observed a general increase in the mean sSFR with increasing redshift. 
Another recent study by \cite{2024ApJ...976...83K} explored the low redshift evolution of SFMS in low-mass galaxies and concluded that there is a significant evolution in the SFMS over the last $\sim$2.5 Gyr (redshift range of 0.05 -- 0.21). A detailed study of a more complete sample of post-merger and non-interacting samples in the local universe will provide more insights on the evolution of sSFR as a function of redshift at lower redshifts.

\section{Summary} \label{6} 
Dwarf galaxies constitute the most prominent population of galaxies across all redshifts. Therefore, the majority of mergers are expected among them, but the impact of mergers among dwarfs is not well explored. In this particular study, we performed ultraviolet (UV) analysis for a sample of 6155 (194 post-merger sample, with tidal features, and 5961 non-interacting) dwarf galaxies within stellar mass range (10$^{7}$ -- 10$^{9.6}$ M$_{\odot}$) and redshift range (0.00 -- 0.12) using FUV imaging from the GALEX mission. This sample is taken from the study of \cite{2020AJ....159..103K}, who identified recent major merger dwarf galaxies using deep optical images from the HSC-SSP. We have estimated the instantaneous SFR for our sample dwarfs using FUV luminosity by developing an automated Python code. The post-merger and the non-interacting sample share a similar SFMS, indicating that the post-merger galaxies are not dominated by either starburst galaxies or quenched galaxies.\\
\indent To find the effect of mergers on star formation, we estimated the difference in log(SFR) between a post-merger galaxy and the median of its corresponding control sample (5 galaxies per one post-merger galaxy, matched in stellar mass and redshift). The offset in our sample has a range ($-$2 to $+$2 dex, about 100 times suppression/enhancement), indicating both enhancement and suppression of star formation in these recent merger galaxies. Around 67\% of the sample (130 galaxies) show an enhancement in SFR. The median offset (enhancement) of the sample is 0.24 dex (1.73 times), indicating a $\sim$ 70\% increase in the SFR of recent merger galaxies compared to their non-interacting counterparts. Out of the 194 post-merger samples, around 44\%, 20\%, and 9\% show 2, 5, and 10 times enhancements in SFR, respectively. On average, we found a moderate enhancement in the median SFR of the post-merger sample, compared to that of the non-interacting control dwarfs, by a factor of nearly two. This factor is similar to the average enhancement factor observed in post-merger massive galaxies. However, we observed widespread star formation across the sample of dwarf galaxies. Star formation is found to be enhanced in both the central (6" diameter region at the centre) and outer regions of the post-merger galaxies compared to their non-interacting counterparts, and the factor of enhancement is found to be similar. This is in contrast to what is observed in massive galaxies, where the merger-triggered star formation is observed to be more significant in the central regions. While investigating the potential dependence of SFR enhancement on stellar mass, no significant variation in enhancement was observed.
Additionally, we found that in a given small range of redshift, post-merger dwarfs exhibit a higher median specific star formation rate compared to their non-interacting counterparts. About 33\% of the galaxies in our post-merger dwarf sample are quenched. These galaxies could be at a later stage of the post-merger regime, where quenching can happen, as observed in massive galaxies. This study suggests that dwarf-dwarf mergers can affect star formation in the local universe. A more detailed study of post-merger dwarfs is required to better understand their impact on galaxy evolution.

\indent This study contains a sample spanning a wide range of stellar mass (10$^{7}$ -- 10$^{9.6}$ M$_{\odot}$) but we did not have enough post-merger dwarfs within the stellar mass range of $10^{7} - 10^{8} M_{\odot}$. Therefore, an extensive range of the lower end of the low mass regime $ 10^{6} M_\odot-10^{8} M_\odot$, at different phases of interactions, still remains unexplored. In the future, we plan to expand our study to include a larger sample of dwarfs with stellar mass ranging from $ 10^{6} M_\odot-10^{8} M_\odot$  and covering various stages of interactions such as mergers, fly-bys, interacting pairs, etc. This will allow us to better understand the impact of these interactions on their star formation. Additionally, future high-resolution UV observations with deeper sensitivity will enhance our understanding of how dwarf-dwarf interactions influence the spatial distribution of star-forming clumps.\\

\begin{acknowledgements}
     It is a pleasure to thank the referee for the insightful and constructive suggestions, which helped to improve the manuscript.  RC thanks Renu and Shashank Gairola for their help during the methodology development stage and Prajwel Joseph for suggestions regarding the automation of the code. This publication utilizes the archival data from GALEX (Galaxy Evolution Explorer), which is a NASA Small Explorer. We extend our gratitude and appreciation to the MAST archive team for making the GALEX data products available online in different formats. RC acknowledges financial support from the Council of Scientific and Industrial Research (09/0890(17755)/2024-EMR-I). SS acknowledges the support from the Science and Engineering Research Board of India through POWER grant(SPG/2021/002672)and support from the Alexander von Humboldt Foundation. MD gratefully acknowledges the support of the SERB Core Research Grant (CRG/2022/004531) and the Department of Science and Technology (DST) grant DST/WIDUSHI-A/PM/2023/25(G) for this research. This work utilized the community-developed core Python packages for Astronomy like Astrodendro {\url{ (https://dendrograms.readthedocs.io/en/stable/)}}, Astropy \citep{2013ascl.soft04002G,Price-Whelan_2018,2022ApJ...935..167A},
      Matplotlib \citep{2007CSE.....9...90H} and Numpy \citep{2020Natur.585..357H} software packages to analyze and process astronomical data.
\end{acknowledgements}

\bibliographystyle{aa} 
\bibliography{reference} 

\appendix 
\onecolumn

\section{SFR Enhancement as a function of stellar mass}

   \begin{figure*}[ht!]
   \centering
   \includegraphics[width=0.49\textwidth, height=0.35\textwidth]{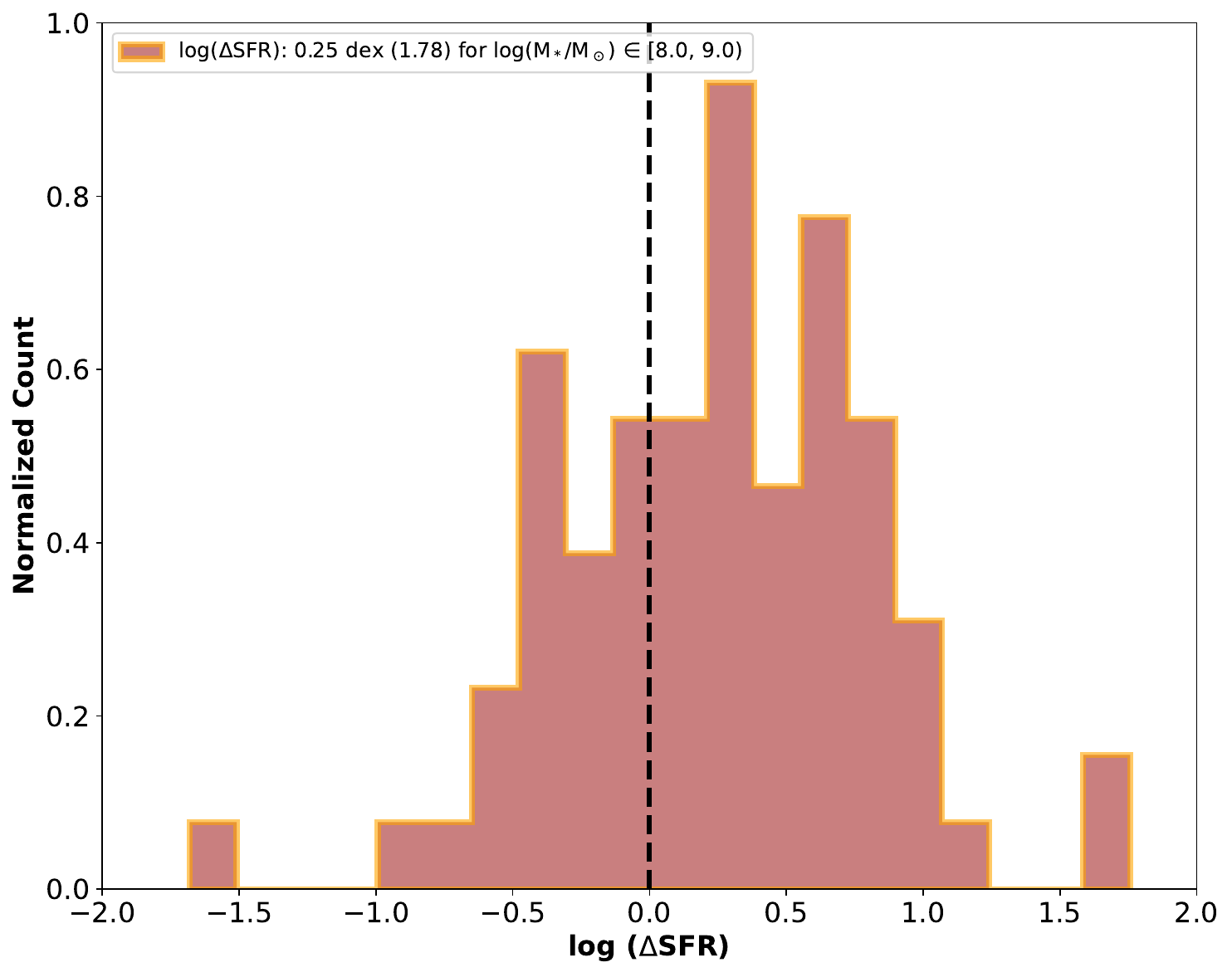}
   \includegraphics[width=0.50\textwidth,height=0.345\textwidth]{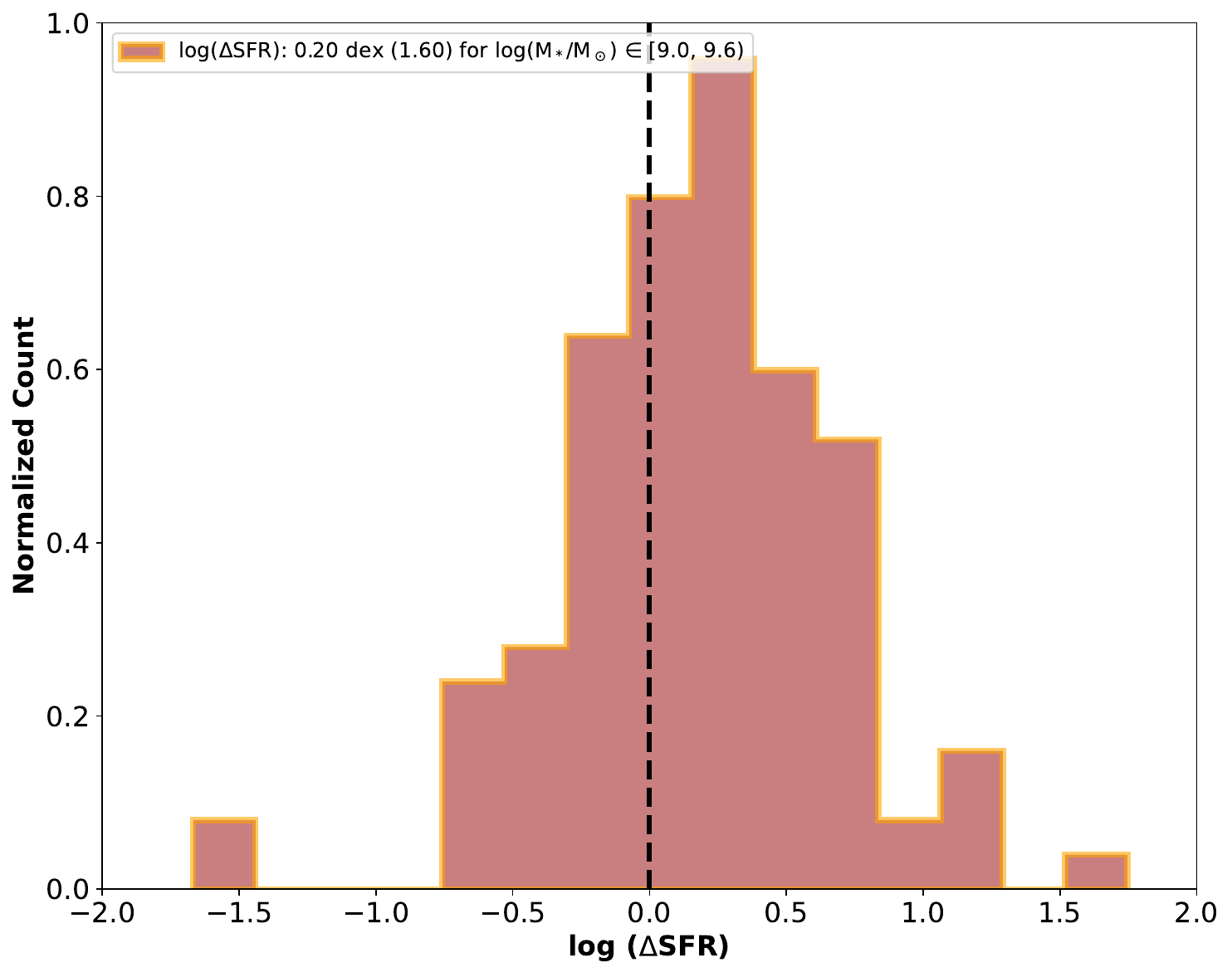}
   \includegraphics[width=0.49\textwidth, height=0.35\textwidth]{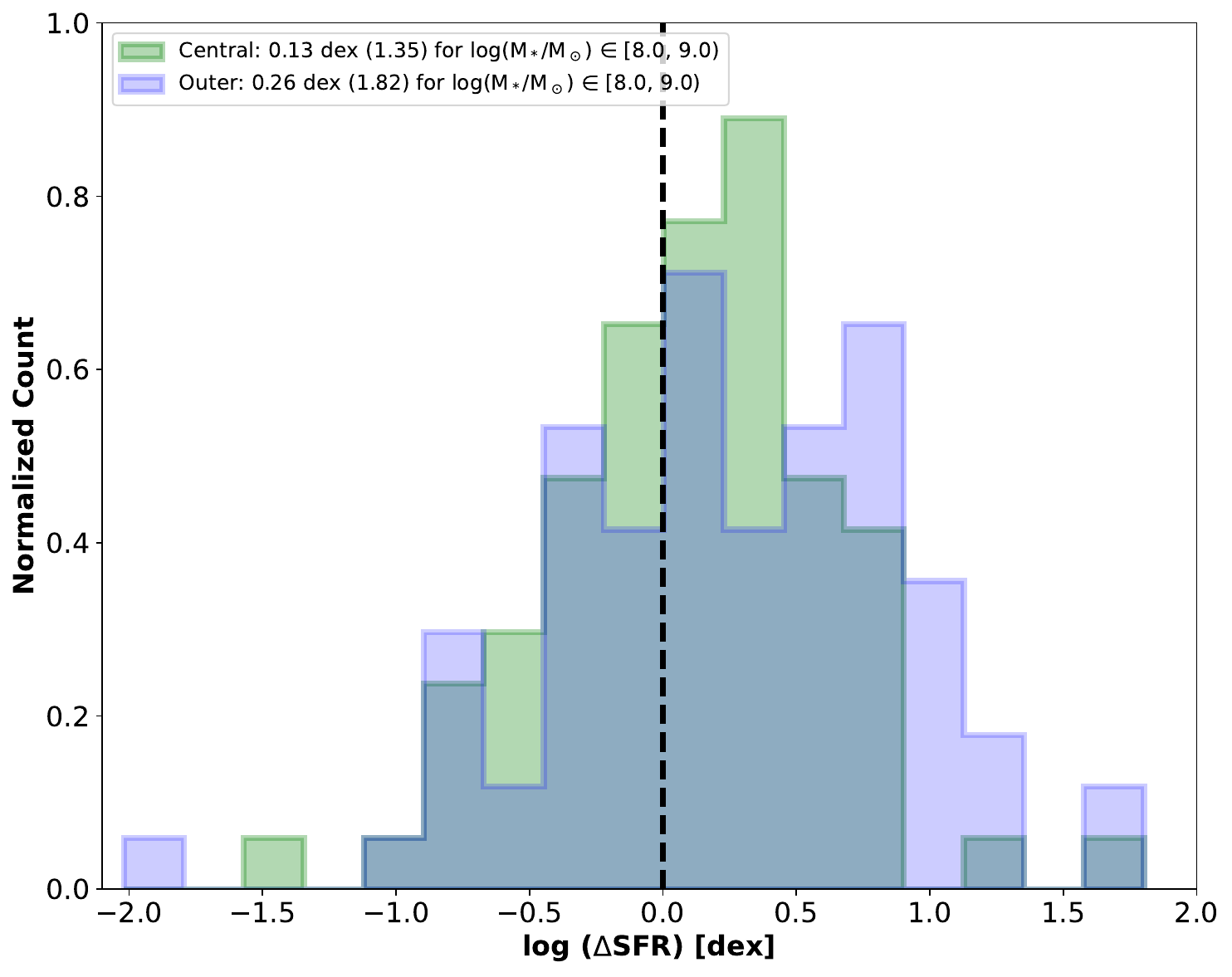}
   \includegraphics[width=0.50\textwidth, height=0.345\textwidth]{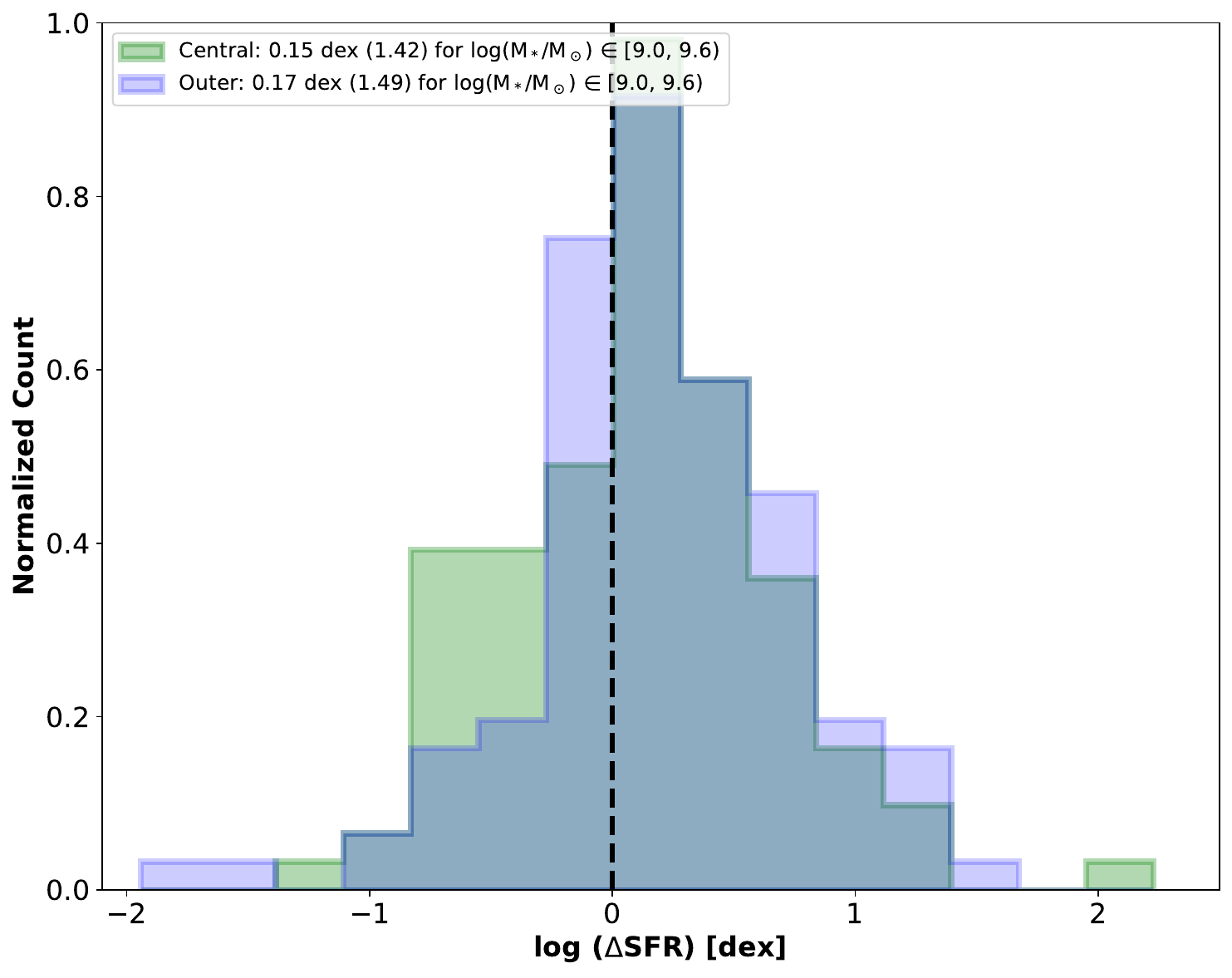}
    \caption{The plots comparing the difference in the SFR (log($\Delta$SFR) =  logSFR$_{post-merger}$ - logSFR$_{non-interacting}$) of the post-merger concerning their non-interacting control dwarfs. Left panel: This figure shows the histogram distribution of SFR offset for the entire galaxy, central, and outer regions for those dwarfs spanning a stellar mass range of $10^{8}-10^{9} M_{\odot}$. Right panel: This figure shows the histogram distribution of SFR offset for the visually identified dwarf galaxy sample, central and outer regions within a stellar mass range of $10^{9}-10^{9.6} M_{\odot}$} 
         \label{Fig9}
   \end{figure*}

\end{document}